\documentclass{nrlreport}

\usepackage[ruled]{algorithm2e} 
\usepackage{listings}
\usepackage{fontenc}
\usepackage{fancyvrb}

\nrlbranch{American Society for Engineering Education Postdoctoral Fellow }
\nrldistribution{A} 
\nrlmark{}
\newreportnumber{7354}{FR}{2022}{2}% formal report %pcm16dec2020 default STRN format
\nrlstartdate{01 OCT 2019}% project start date
\nrlenddate{30 SEP 2022}% project end date
\nrlprogramelement{0601153N}
\nrlprojectnumber{100001764031.0010}
\nrlworkunitnumber{1R32}

\nrlorganizationaddress{US Naval Research Laboratory \\1005 Balch Blvd. \\Stennis Space Center, MS 39529}
\nrlreportclass{U/U}
\nrlabstractclass{U/U}
\nrlthispageclass{U/U}
\nrlabstractlimitation{SAR}
\nrlresponsibleperson{Allison M. Penko}
\nrltelephonenumber{(228) 688-4823}

\nrltitlepagedate{July 28, 2022}% the date that goes on the title page %version
\nrlreportdate{07-28-2022}% the date the goes on the documentation page
\author{William S. Kearney}
\nrlheaderauthor{Kearney and Penko}
\date{\today}
\title{The Naval Seafloor Evolution Architecture: A Platform for Predicting Dynamic Seafloor Roughness}
\nrlheadertitle{NSEA: Predicting seafloor roughness}
\hypersetup{
 pdfauthor={William S. Kearney, Allison M. Penko},
 pdftitle={The Naval Seafloor Evolution Architecture: a platform for forecasting dynamic seafloor roughness},
 pdfkeywords={},
 pdfsubject={},
 pdfcreator={Emacs 26.3 (Org mode 9.1.9)}, 
 pdflang={English}}

\nrlauthortwo{Allison M. Penko}
\nrlbranchtwo{Seafloor Sciences Branch\\Ocean Sciences Division}

\begin{document}

\maketitle
\nrlabstract{
Predicting the temporal and spatial dynamics of seafloor roughness is important for understanding bottom boundary layer hydrodynamics. The Navy Seafloor Evolution Architecture (NSEA) is a platform for modeling the dynamic nature of the seafloor by combining hydrodynamic forcing information and observations from diverse sources. NSEA's three modules include a specification of hydrodynamic forcing, a seafloor evolution model, and a model to generate roughness realizations. It can be run in forward mode to predict seafloor roughness including the uncertainty from forcing information, or in inverse mode to estimate parameters from observed seafloor roughness. The model is demonstrated and shown to have good agreement with a field dataset of observed seafloor roughness. Similarly running in inverse mode, NSEA was demonstrated to predict the observed mean sediment grain size with good agreement. NSEA's modularity allows for a wide range of applications in hydrodynamic and acoustic modeling, and is built within an expandable framework that lends for coupling to such models with minimal effort. 
}

\tableofcontents

\chapter{Introduction}
\label{sec:orgb604d24}

The vertical and horizontal variation of the seafloor impacts ocean wave attenuation, acoustic propagation, and the scour and burial of objects on the seabed. These elevation variations cover a wide range of spatial and temporal scales, from the grain scales of seafloor sediments (\(\mathcal{O}(100\units{mm})\)) to the ocean basins (\(\mathcal{O}(10,000 \units{km})\)). However, bathymetric data is available only at a finite spatial and temporal resolution. The General Bathymetric Chart of the Ocean's global gridded bathymetry data, for example, is provided at a 15-arc-second resolution, which is approximately 450 meters at the equator. "Roughness" refers to the variability in seafloor elevation below this grid scale, though the exact delineation between roughness and bathymetry varies with the available data. 

Seafloor roughness is particularly important to hydrodynamics, because it enhances turbulence within the bottom boundary layer and exerts friction on the overlying flow \cite{Mathisen_1996,Styles_2002}, and acoustics, because it modifies the scattering of high-frequency sound waves from the sediment-water interface \cite{Jackson_2007}. Roughness also evolves as sediment is transported and bedforms such as ripples and dunes are created or destroyed. Because roughness is not measured by definition, it must be predicted from observed or modeled geological properties and hydrodynamic forcing. High-resolution numerical models \cite{Marieu_2008,Penko_2013,Mazzuoli_2019,Salimi-Tarazouj_2021a,Salimi-Tarazouj_2021b} can directly simulate the interaction of hydrodynamics and sediment transport in the bottom boundary layer, resulting in the formation and evolution of bedforms. However, these models are computationally intensive and are not suitable for embedding in operational hydrodynamic or acoustic models. Embedded models are statistical in nature, predicting an estimate of the roughness such as a length scale or power spectrum. They typically rely on empirical parameterizations developed from data collected in laboratory and field settings. The majority of models used operationally are equilibrium models that predict ripple heights and lengths from instantaneous hydrodynamic information \cite{Nielsen_1981,VanRijn_1984c,Wiberg_1994,Nelson_2013,Goldstein_2013}. These models are computationally efficient, but they fail to predict roughness out of equilibrium with the forcing. Non-equilibrium phenomena are most evident during ripple formation, as it takes time for ripples to adjust to new forcing \cite{Traykovski_2007}, and under relict conditions, in which ripples persist under hydrodynamic forcing not strong enough to mobilize sediment \cite{Hay_2008,DuVal_2021}.

Non-equilibrum roughness models \cite{Traykovski_2007,Soulsby_2012,Nelson_2015,Penko_2017} simulate the time evolution of roughness under changing forcing conditions and have been used successfully to predict roughness on shallow continental shelves. These models are statistical models, typically predicting roughness power spectra rather than the exact evolution of the seafloor elevation. They rely on empirical parameterizations of both sediment transport \cite{MeyerPeter_1948} and equilibrium ripple geometry, but embed these empirical parameterizations within a physics-based model for evolution.

The Naval Seafloor Evolution Architecture (NSEA) evolved from a non-equilibrium roughness model also known as NSEA (the Navy Seafloor Evolution Archetype \cite{Penko_2017}).\footnote{Further uses of NSEA within this technical report refer to the Naval Seafloor Evolution Architecture, and the non-equilibrium roughness model derived from \cite{Penko_2017} will be called the Navy Seafloor Evolution Model (NSEM).} This roughness model now forms the central part of a larger framework for characterizing the spatial and temporal evolution of seafloor roughness on operationally relevant time scales. This technical report describes in detail the components of the framework in Section \ref{sec:orgdc16ebd}, then discusses the application of statistical inference methods within the model in Section \ref{sec:orga29e677}. Section \ref{sec:orgc193e3b} describes a reference implementation of NSEA in the Julia programming language \cite{Bezanson_2017}, and Section \ref{sec:org2def406} applies NSEA to seafloor roughness prediction and the estimation of sediment properties using data from a field experiment. Finally, Section \ref{sec:orgf3071a0} summarizes the current state of the architecture and presents future directions for research and development of NSEA.

\section{Statistical Characterization of Seafloor Roughness}
\label{sec:orgbc87f4e}
Before describing NSEA, it is valuable to describe how to characterize the statistical properties of the seafloor elevation and introducing notation that will be used throughout this technical report. The elevation of the seafloor, \(\eta(\mathbf{x},t)\), is a spatiotemporal \emph{random field}, a stochastic process that is indexed by space, \(\mathbf{x} \in \mathbb{R}^2\), and time, \(t \in \mathbb{R}^+\). The issues that arise when modeling the seafloor elevation as a stochastic process will be discussed below, but for now, the seafloor elevation field is assumed to be described by a probability distribution that gives the likelihood of observing a particular configuration of the seafloor elevation in space and time. The goal of this analysis is to compute certain statistical characteristics of the random seafloor elevation. These include the mean function

\begin{equation}
\mu(\mathbf{x},t) = \langle \eta(\mathbf{x},t) \rangle,
\end{equation}

\noindent the covariance function

\begin{equation}
K(\mathbf{x}_1,t_1,\mathbf{x}_2,t_2) = \langle \left(\eta(\mathbf{x}_1,t_1) -\mu(\mathbf{x}_1,t_1)\right) \left(\eta(\mathbf{x}_2,t_2) - \mu(\mathbf{x}_2,t_2)\right) \rangle,
\end{equation}

\noindent and higher-order statistics of the form

\begin{equation}
R_n(\mathbf{x}_1,t_1,\dots,\mathbf{x}_n,t_n) = \langle \eta(\mathbf{x}_1,t_1) \dots \eta(\mathbf{x}_2,t_2) \rangle
\end{equation}

\noindent that are called the "moments" of the random field. 

\newpage

Other important statistical properties can be derived from these moments, such as the mean square roughness:

\begin{equation}
\langle \eta(\mathbf{x},t)^2 \rangle = K(\mathbf{x},t,\mathbf{x},t).
\end{equation}

The operator \(\langle \cdot \rangle\) represents the average (expectation) under the probability distribution for the elevation field. This is an \emph{ensemble average}, and it does not correspond to a spatial or temporal average. As a result, the moments all generally depend on both space and time.

NSEA assumes that the seafloor elevation is wide-sense stationarity in space \cite{Guttorp_1995}. A wide-sense stationary process has a constant mean function (\(\mu(\mathbf{x},t) = \mu(t)\)), and the equal-time covariance function \(K_t(\mathbf{x}_1,\mathbf{x}_2) = K(\mathbf{x}_1,t,\mathbf{x}_2,t)\) depends only on the difference between the points \(\mathbf{x}_1\) and \(\mathbf{x}_2\): \(K_t(\mathbf{x}_1,\mathbf{x}_1 + \mathbf{h}) = K_t(\mathbf{h})\). Stationarity does not hold in the real world, and the statistical properties of the seafloor elevation field will vary spatially as parameters like the hydrodynamic forcing, water depth, and geological properties of the seafloor change \cite{Holland_2008}. The assumption of stationarity in NSEA limits us to modeling patches of seafloor over which these parameters do not change significantly.

A further simplification results from assuming that the mean elevation within the patch of interest is zero. Since the seafloor elevation is always measured relative to an arbitrary vertical datum, this does not result in any loss of information. The covariance function \(K_t(\mathbf{h})\) is then equivalent to the second moment, \(R_2\).

A wide-sense stationary process has a translation invariant covariance function, but is not completely characterized by the covariance function. Higher-order moments can be important in generating realistic seafloor statistics \cite{Tang_2009b}, but NSEA does not explicitly model higher-order moments, and in practice, they are assumed to be zero. The seafloor elevation is thus modeled as a zero-mean Gaussian process, which is completely characterized by its covariance function, \(K_t(\mathbf{h})\). A Gaussian process has the property that all the finite-dimensional marginal distributions are multivariate Gaussian \cite{GPML,Guttorp_1995}. This means that if the seafloor elevation is measured at a set of points \(\{\mathbf{x}_i\}_{i=1}^N\), the distribution of the vector \([\eta(\mathbf{x}_1,t) \dots \eta(\mathbf{x}_N,t)]\) is a multivariate Gaussian with mean zero and covariance matrix

\begin{equation}
\Sigma_{ij} = K_t(\mathbf{x}_i - \mathbf{x}_j).
\end{equation}

Stationary processes can alternatively be characterized by the Fourier transform of the covariance function

\begin{equation}
S_t(\mathbf{k}) = \frac{1}{2\pi} \int K_t(\mathbf{h}) e^{-i\mathbf{k} \cdot \mathbf{h}}\ d\mathbf{h},
\end{equation}

\noindent which is called the \emph{spectral density}, or \emph{power spectrum}, of the process \cite{GPML}. The spectral density for a two-dimensional random field has dimensions of \(\mathrm{length}^4\) because the covariance function and the area element, \(d\mathbf{h}\), both have dimensions of \(\mathrm{length}^2\). The spectral density measures the average squared amplitude of fluctuations in the seafloor elevation that have a spatial frequency given by the wavenumber, \(\mathbf{k}\). The integral of the spectral density over all wavenumbers is the mean-square surface roughness

\begin{equation}
\langle \eta(\mathbf{x},t)^2 \rangle = K(\mathbf{0}) = \int S_t(\mathbf{k})\ d\mathbf{k}.
\end{equation}

The covariance function and the power spectrum are equivalent in the sense that they can be converted between each other using the Fourier transform. In spatial statistics, it is more common to model the covariance function by selecting from families of well-understood covariances \cite{GPML} or by developing models of spatial correlation between neighboring sites such as Markov random fields \cite{Lindgren_2011}. These approaches most often make the further assumption of isotropy, that the covariance function depends only on the distance, \(\|\mathbf{h}\|\), between the measurement points. The resulting random field is rotation- and translation-invariant. Rippled seafloors are clearly anisotropic, with a preferred direction, and oscillations that are not accurately characterized by isotropic covariance functions. It proves somewhat easier to represent anisotropic and oscillating covariances in the spectral domain, which is one reason NSEA models the power spectrum rather than the covariance function. The other major reason to prefer the spectral representation is that valid covariance functions must be positive definite, and this constraint is hard to ensure while evolving the covariance function in time. However, as long as the power spectrum is positive, the corresponding covariance function is necessarily positive definite via Bochner's theorem \cite{GPML}. NSEA can evolve the power spectrum forward in time in such a way that it is always greater than zero, and the resulting covariance function is always valid.

The covariance function and power spectrum presented above are computed at a single time. There are equivalent spatiotemporal covariance functions and spectra that characterize the correlation of the seafloor elevation at unequal times. However, modeling these explicitly introduces additional complexity \cite{Gneiting_2007}, and the spatiotemporal statistical properties of real seafloors are not well understood. NSEA ignores the temporal correlations in the seafloor dynamics, assuming that observations of the seafloor separated by times much longer than the characteristic time scales for the forcing, such as the wave period, will be uncorrelated.

Modeling the seafloor elevation as a stochastic process implies that there is some source of randomness in this system. However, the evolution of seafloor roughness is governed by deterministic physical processes of hydrodynamics and sediment transport. A natural question is what the probability distribution over the seafloor elevation actually represents. Deterministic physics are not resolved at the spatial and temporal resolutions necessary for a complete, deterministic prescription. Forcing information is necessarily averaged over some spatial and temporal scales. For example, NSEA might be driven with a wave model such as SWAN \cite{SWAN_2021}. The output of SWAN is the power spectrum of the sea surface height within a grid cell of the model. Like with NSEA, the actual spatiotemporal evolution of the sea surface height is replaced by a statistical description. There are therefore many potential realizations of the sea surface height within a grid cell consistent with the modeled wave spectrum, and each such realization would drive the evolution of the seafloor in a slightly different way. A statistical description of the forcing therefore requires a statistical description of the seafloor evolution. This is especially true if the seafloor evolution model is forced by reduced statistics of the forcing such as the significant wave height and period, which contain even less information about the detailed hydrodynamic forcing than the power spectrum.

Furthermore, even if one could completely resolve the sea surface height, waves drive sediment transport through a turbulent bottom boundary layer that exhibits deterministic but chaotic dynamics \cite{Leith_1996}. Small perturbations in the flow within the bottom boundary layer can lead to very different patterns of sediment transport even under a fully specified wave forcing. The statistical description of seafloor evolution employed by NSEA lumps the uncertainty in the chaotic bottom boundary layer dynamics in with the uncertainty in the forcing. The randomness in the seafloor elevation field ultimately comes from the influence of these unresolved processes. The averaging operator \(\langle \cdot \rangle\) used in the definition of the moments of the seafloor elevation field above represents an average over the random forcing and the chaotic turbulent dynamics.

\chapter{System architecture}
\label{sec:orgdc16ebd}
NSEA consists of four modules: forcing, seafloor evolution, seafloor synthesis, and observations. Information naturally flows from one module to the next, so that the seafloor evolution module consumes forcing information and its outputs are, in turn, consumed by the seafloor synthesis module. However, NSEA also enables information to flow in the reverse direction in two ways. First, because NSEA represents the evolution of the seafloor in time, outputs from one module can be fed back into a previous model. The typical use case for this kind of coupling in time would be to use the roughness information output from the seafloor evolution model to parameterize bottom friction in a hydrodynamic model that supplies the forcing. Each module's inputs and outputs are directly accessible to the user to facilitate feedbacks of this nature. The second way to share information in the reverse direction is to use techniques of statistical inference to estimate parameters and state variables of each module from available data. For example, observations of bedforms can constrain the seafloor elevation field or its power spectrum, which are normally simulated by the seafloor synthesis and evolution modules. NSEA requires that each module be constructed as a probabilistic model to enable these techniques, which are discussed in more detail in Section \ref{sec:orga29e677}.

\section{Forcing}
\label{sec:org4eb48a9}
The forcing information required by NSEA consists of time series of wave and/or current forcing. Wave forcing consists of a bottom orbital velocity, \(u_w [\units{m \cdot s^{-1}}]\), a bottom semi-orbital excursion, \(A_w [\units{m}]\), and a wave direction, \(\varphi_w [\units{rad}]\), while current forcing consists of a depth-averaged current velocity, \(u_c [\units{m \cdot s^{-1}}]\), and a current direction, \(\varphi_c [\units{rad}]\). These time series can be sampled at any interval including irregular sampling intervals, but the seafloor evolution model assumes that the forcing is constant over the sampling intervals. It is important to provide forcing information at a high enough resolution to capture the desired scales of temporal variability. NSEA is agnostic as to the source of the forcing information, as long as it is supplied as a time series of the required forcing variables. Forcing data can therefore come from in situ or remotely sensed observations or from hydrodynamic model output. 

The predictions and statistical inferences about the seafloor roughness that NSEA makes are always conditional on the supplied forcing information. If the user wishes to propagate uncertainty in the forcing through the model or integrate over the forcing uncertainty in performing statistical inferences, they must run ensemble simulations of NSEA in which each ensemble member is driven by a different forcing.

Wave bottom orbital velocities and excursions are often not directly observed in the field. They can be calculated from surface wave statistics by using linear wave theory. For a monochromatic wave with amplitude \(H\) and period \(T\) in a water depth \(h\), the wave bottom orbital velocity is \cite{Wiberg_2008}

\begin{equation}
\label{eq:org8ea0090}
u = \frac{H \pi}{T \sinh kh},
\end{equation}

\noindent and the wave semi-orbital excursion is given by \(A_w = u T / 2\pi\). The wavenumber \(k\) is determined from the dispersion relation

\begin{equation}
\frac{4\pi^2}{T^2} = gk \tanh kh,
\end{equation}

\noindent which must be solved iteratively for \(k\) \cite{Wiberg_2008}. 

For a continuous spectrum of surface waves, there is some ambiguity in the choice of wave bottom orbital velocity. The root-mean-square wave bottom orbital velocity is given by \cite{SWAN_2021}

\begin{equation}
u_{RMS}^2 = \int_0^{2\pi}\int_0^{\infty} \frac{\sigma^2}{sinh^2 kh} E(\sigma,\theta)\ \mathrm{d}\sigma\mathrm{d}\theta,
\end{equation}

\noindent where \(E(\sigma,\theta)\) is the directional wave power spectral density with relative frequency, \(\sigma\), and direction, \(\theta\). For a monochromatic wave, \(u = \sqrt{2}u_{RMS}\). NSEA expects wave bottom orbital velocity to be provided as \(u_{1/3}\), the velocity derived from \eqn{eq:org8ea0090} with a height equal to the significant wave height, but it does not enforce this condition, and it is up to the user to provide the appropriate wave bottom orbital velocities and excursions.

\section{Seafloor Evolution}
\label{sec:org5ad68e6}
The seafloor evolution model, \cite{Traykovski_2007,Nelson_2015,Penko_2017}, approximates the evolution of the amplitude spectrum of the seafloor elevation (\(\sqrt{S}\ [\units{m^2}]\)):

\begin{equation}
\label{eq:orgbbdddb1}
\frac{\partial \sqrt{S_t(k)}}{\partial t} = \frac{1}{T_t(k)}\left(\sqrt{\bar{S}_t(k)} - \sqrt{S_t(k)}\right).
\end{equation}

Equation (\ref{eq:orgbbdddb1}) models the relaxation of the amplitude spectrum toward its equilibrium value, \(\sqrt{\bar{S}_t(k)}\), with a time scale of \(T_t(k)\). The forcing information enters the seafloor evolution model through the relaxation time scale and the equilibrium ripple spectrum. This model can be interpreted as a linearization of a nonlinear evolution equation (e.g., \cite{Davis_2004}) around the equilibrium ripple spectrum.

Traykovski \cite{Traykovski_2007} derives the relaxation time scale from the sediment continuity equation, leading to the result that the relaxation time scale is proportional to the cross-sectional area of the ripple divided by the width-averaged sediment transport rate. For a sinusoidal ripple with a height of \(\eta\) and a wavelength of \(\lambda\), the cross-sectional area can be approximated by \(A = \eta \lambda/2\). The width-averaged volumetric sediment transport rate, \(\hat{Q}\), is scaled by the bed packing density, \((1-\phi)\), where \(\phi\) is the porosity, leading to the relationship

\begin{equation}
T(k) \propto \left(1 - \phi\right) \frac{\lambda \eta}{2\hat{Q}} = \frac{1 - \phi}{2\hat{Q}} \alpha \left(\frac{2\pi}{k}\right)^2,
\end{equation}

\noindent where in the second equality, the ripple height is related to the wavelength by a constant steepness, \(\eta = \alpha \lambda\), and \(\lambda\) is rewritten in terms of the ripple wavenumber, \(\lambda = 2\pi/k\). The parameter \(\alpha\) represents the typical steepness of ripples. Traykovski gives \(\alpha = 0.16\), which is generally found to be the case for wave-orbital ripples \cite{Wiberg_1994}. However, \(\alpha\) can be combined with the unknown proportionality constant and used as a tuning parameter \cite{Penko_2017}.

The model represents the peak sediment transport rate, \(\hat{Q}\), as a function of the hydrodynamic forcing. One could use any empirical sediment transport law or a more complex bottom boundary layer model to determine \(\hat{Q}\). The default transport formula that is used in NSEA is the bedload transport law of Meyer-Peter and M\"{u}ller \cite{MeyerPeter_1948}:

\begin{equation}
\label{eq:org0795cab}
\hat{Q} = \gamma\left(\theta - \theta_{cr}\right)_{+}^{\xi} / \left(\rho (s - 1) g D_{50}\right),
\end{equation}

\noindent where \(\theta\) is the non-dimensionalized shear stress, \(\theta_{cr}\) is the non-dimensionalized critical shear stress, and the notation \((x)_{+} = \max(x,0)\). The empirical parameters \(\gamma\) and \(\xi\) are typically assigned values of 8 and 1.5 in the literature \cite{Wong_2006}.

The equilibrium ripple spectrum, \(\sqrt{\bar{S}_t(k)}\), represents the ripple geometry to which the seafloor would evolve under a constant forcing. Following \cite{Traykovski_2007,Soulsby_2012,Nelson_2015,Penko_2017}, the equilibrium power spectrum is modeled as a Gaussian function centered on an equilibrium wavenumber, \(\bar{\mathbf{k}}\), and with an amplitude proportional to the square of the equilibrium ripple height, \(\bar{\eta}\):

\begin{equation}
\label{eq:org5019909}
S_t(\mathbf{k}) = \frac{\bar{\eta}^2}{16\pi\sigma^2} e^{-\frac{\|\mathbf{k} - \bar{\mathbf{k}}\|^2}{2\sigma^2}}.
\end{equation}

The width of the spectrum around the equilibrium wavelength is a parameter \(\sigma\). This parameter could also vary separately in the x and y directions \cite{Nelson_2015} with appropriate modifications to \eqn{eq:org5019909}. The spectrum is scaled so that the mean squared ripple amplitude is \(\bar{\eta}^2/8\), so that as the spectral width tends to zero, the resulting spectrum models a sinusoidal ripple field with height \(\bar{\eta}\).

The equilibrium ripple height and the wavenumber (or, equivalently, the wavelength, \(\bar{\lambda} = 2\pi/\|\mathbf{k}\|\)) are obtained by an empirical equilibrium ripple geometry formula \cite{Nielsen_1981,Wiberg_1994,Nelson_2013,Soulsby_2012,Goldstein_2013} that predicts the equilibrium ripple geometry as a function of the instantaneous hydrodynamic forcing. NSEA can use a wide variety of ripple predictors, which enables testing of different parameterizations. An additional ripple predictor is included to allow testing the sensitivity of the model to ripple geometry parameterizations. It takes the form of a power law in the mobility number \(\psi = u_w^2 / g(S-1)D_{50}\) and the displacement parameter \(\delta = A_w / D_{50}\):

\begin{equation}
\label{eq:org13961b3}
\frac{\bar{\lambda}}{A_w} = e^{\beta_1}\psi^{\beta_2}\delta^{\beta_3} 
\end{equation}

\begin{equation}
\label{eq:org13961b3a}
\frac{\bar{\eta}}{\bar{\lambda}} = e^{\chi_1}\psi^{\chi_2}\delta^{\chi_3}. 
\end{equation}

The exponents \(\{\beta_i\}_{i=1}^3\) and \(\{\chi_i\}_{i=1}^3\) are free parameters. Many existing ripple predictors have similar forms to this power law or can be approximated fairly well using the power law (Fig. \ref{fig:org6bf012a}).

\href{figures/ripple\_predictors.png}{\begin{figure}
\centering
\includegraphics[width=.9\linewidth]{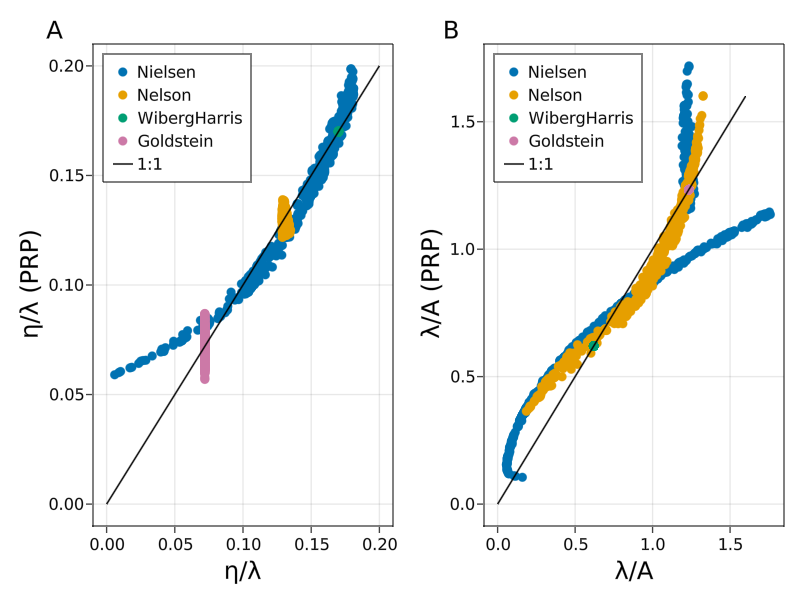}
\caption{\label{fig:org6bf012a}Comparison between the non-dimensionalized ripple height (A) and wavelength (B) for the ripple predictors from the literature and the power law model. The x-axis in each plot shows the ripple geometry predicted by each formula while the y-axis is the geometry given by the power law fitted to the predictions of each ripple formula. Ideally, the predictions would fall along the black 1:1 line.}
\end{figure}}

Under strong forcing conditions, sheet flow conditions are expected to exist, in which case the entire bed is mobilized and any ripples present are washed out. This is accomplished by setting the ripple height to zero and the wavelength to infinity when the nondimensional shear stress is above the washout threshold, \(\theta_{wo}\). This is accomplished within the model by modifying the equilibrium ripple height and wavelength when above the washout threshold and setting the adjustment time scale to a small number (typically 60 s) so that the spectrum evolves normally, but rapidly decays toward a flat bed.

In the absence of hydrodynamic forcing, it is expected that bioturbation will slowly modify the roughness, leading to a decay of ripples \cite{Jackson_2009}. This is modeled as a diffusive process by adding a term to \eqn{eq:orgbbdddb1}:

\begin{equation}
\label{eq:orga550ad4}
\frac{\partial \sqrt{S_t(k)}}{\partial t} = \frac{1}{T_t(k)}\left(\sqrt{\bar{S}_t(k)} - \sqrt{S_t(k)}\right) - Dk^2 \sqrt{S_t(k)}.
\end{equation}

The diffusion process is always on, regardless of the forcing. Since the diffusion coefficient is typically on the order of \(10^{-9} \units{m^2 \cdot{} s^{-1}}\) \cite{Jackson_2009, Penko_2017, Penko_2015}, it has little impact on the ripple geometry when sediment is actively being transported, but results in a slow decay of ripples during quiescent periods.

\subsection{Alternative Derivation}
\label{sec:orgbf5c576}
The spectral evolution equation models the evolution of the amplitude spectrum over a patch of seafloor, not the fine-scale dynamics of the seafloor elevation itself. One weakness of this approach is the difficulty in extending the spectral evolution model to include additional forcings such as combined waves and currents, internal waves, or additional sedimentological processes such as sediment cohesion and bioturbation. Furthermore, the model posed in \eqn{eq:orgbbdddb1} is ultimately derived heuristically rather than through a physical model for the flux itself.

Therefore, it is invaluable to explore how fine-scale models for sediment transport lead to statistical roughness models like \eqn{eq:orgbbdddb1}. It is not generally possible to derive the statistical roughness model corresponding to a particular fine-scale sediment transport model. However, there is one fine-scale model, a stochastic heat equation, that is analytically tractable and that provides some insight into the non-equilibrium spectral evolution equation.

The stochastic heat equation is given by

\begin{equation}
\label{eq:org013d09e}
\frac{\partial \eta}{\partial t} = D \frac{\partial^2 \eta}{\partial x^2} + \zeta(x,t).
\end{equation}

This equation is written as a one-dimensional process for notational simplicity, but the extension to two dimensions is straightforward. Here, \(\zeta(x,t)\) is a Gaussian process with a mean of zero and a correlation function, \(E[\zeta(x,t)\zeta(x\prime,t\prime)] = C(x-x\prime)\delta(t-t\prime)\). In other words, \(\zeta\) is white noise in time with a stationary spatial correlation function \(C(\cdot)\). This stochastic partial differential equation can be discretized over a patch of length \(L\) with periodic boundary conditions by a Fourier Galerkin method, expanding \(\eta(x,t) = \sum_{k}\eta_k(t)e^{-i2\pi kx/L}\) and \(\zeta(x,t) = \sum_{k}\zeta_k(t) e^{-i2\pi kx/L}\). Substituting these expressions, multiplying by \(e^{i2\pi kx/L}\), and integrating over the patch, the evolution equation becomes, when written in the customary notation for stochastic differential equations,

\begin{equation}
d\eta_k = -Dk^2 \eta_k dt + \sigma_k dW_k,
\end{equation}

\noindent where the variance term, \(\sigma_k^2\), is the Fourier transform of the spatial covariance function, \(C\):

\begin{equation}
\sigma_k^2 = \int C(x) e^{i 2\pi kx/L}\ \mathrm{d}x.
\end{equation}

The noise term, \(W_k\), is a complex circular Brownian process (i.e., \(W_k = U_k/\sqrt{2} + iV_k/\sqrt{2}\)) with \(U_k\) and \(V_k\) as independent real-valued Brownian motions. The Fourier coefficients, \(\eta_k\), are complex variables; however, the real and imaginary parts evolve independently because the coefficients \(\alpha_k = -Dk^2\) and \(\sigma_k\) are real, and the complex white noise term consists of independently sampled real and imaginary components. The complex Fourier coefficients, \(\eta_k = a_k + ib_k\), and \(a_k\) and \(b_k\) follow the stochastic differential equations

\begin{align}
da_k &= -\alpha_k a_k dt + \frac{\sigma_k}{\sqrt{2}} dU_k \\
db_k &= -\alpha_k b_k dt + \frac{\sigma_k}{\sqrt{2}} dV_k.
\end{align}

The power spectrum of the seafloor elevation in this case is given by the expected value of the squared magnitude of the Fourier coefficients, \(S_k = E[|\eta_k|^2] = E[a_k^2 + b_k^2]\). Performing a change of variables using It\^{o}'s formula \cite{Oksendal_2003} gives the stochastic differential equation satisfied by \(c_k = a_k^2 + b_k^2\),

\begin{equation}
dc_k = 2\alpha_k \left(\frac{\sigma_k^2}{2\alpha_k} - c_k\right) dt + \sigma_k\sqrt{2c_k}d\tilde{W}_k.
\label{eq:CIR-mod}
\end{equation}

Equation \eqref{eq:CIR-mod} is a version of the analytically tractable Cox-Ingersoll-Ross model \cite{Cox_1985}, and a deterministic evolution equation given \(E[c_k] = S_k\) is 

\begin{equation}
\frac{dS_k}{dt} = 2\alpha_k \left(\frac{\sigma_k^2}{2\alpha_k} - S_k\right).
\end{equation}

Identifying \(\bar{S}_k = \sigma_k^2 / 2\alpha_k\) and \(T_k = 1/2\alpha_k\), the spectral evolution implied by the stochastic heat equation has a form similar to that of NSEA:

\begin{equation}
\label{eq:orga644bff}
\frac{dS_k}{dt} = \frac{1}{T_k} \left(\bar{S}_k - S_k\right);
\end{equation}

\noindent however, it is the power spectrum that evolves according to a linear relaxation process rather than the amplitude spectrum.

The time scale parameter is related to \(\alpha_k\):

\begin{equation}
\frac{1}{2\alpha_k} = \frac{1}{2Dk^2} = T_k = \frac{1 - \varphi}{2\hat{Q}} \alpha \left(\frac{2\pi}{k}\right)^2,
\end{equation}

\noindent and one can observe that the \(k^2\) dependency of \(\alpha_k\) matches the \(k^{-2}\) dependency of the relaxation time scale according to Traykovski's derivation. The diffusion coefficient, \(D\), can be described as a function of the scaled volumetric sediment transport rate:

\begin{equation}
D = \frac{\hat{Q}}{4\pi^2 \alpha (1 - \varphi)}.
\end{equation}

The equilibrium spectrum, which is obtained through the equilibrium ripple predictor, can easily be converted into the forcing variance, \(\sigma_k^2 = 2\bar{S}_k \alpha_k\). 

This derivation illustrates that a very simple fine-scale sediment transport model consisting of a diffusive relaxation and a random flux of sediment with a spatial pattern determined by the wave forcing leads to a statistical roughness model with dynamics qualitatively similar to that used in NSEA. However, it does suggest that the power spectrum should be evolved forward in time according to \eqn{eq:orga644bff} rather than the amplitude spectrum as in \eqn{eq:orgbbdddb1}. Both of these formulations are models of complex sediment dynamics, and it is not possible to determine from first principles whether the heuristic scaling with the amplitude spectrum used to develop \eqn{eq:orgbbdddb1} or the stochastic sediment flux model used to develop \eqn{eq:orga644bff} is more appropriate. At present, NSEA uses the amplitude spectrum equation \eqn{eq:orgbbdddb1}. Further work is needed to compare the predictions of these and other statistical roughness models to field and laboratory data, and the modular nature of NSEA allows different evolution models to be substituted for one another quite easily.

\section{Seafloor Synthesis}
\label{sec:orge5b6faa}
The output of the seafloor evolution model is the amplitude spectrum, which characterizes the statistical properties of the random seafloor elevation field (\secn{sec:orgbc87f4e}). Useful statistics of seafloor roughness such as the root-mean-square elevation or the peak ripple wavelength can be computed from the spectrum. Many applications, such as estimating bottom friction, use these statistics directly. However, other applications, such as acoustic modeling of seafloor scattering (e.g., \cite{Thorsos_1988,Olson_2020}), require realizations of the random seafloor elevation field. The seafloor synthesis component of NSEA provides these realizations through a Fourier synthesis approach \cite{Thorsos_1988,Paciorek_2007}.

The realizations cover a two-dimensional patch with side lengths \((L_x,L_y)\). The assumption of periodic boundary conditions enables the representation of any seafloor elevation field with a Fourier series. The Fourier decomposition of the seafloor elevation is

\begin{equation}
\label{eq:orgf4173d0}
\eta(\mathbf{x},t) = \sum_{m,n} \eta_{m,n}(t) e^{i\left(2\pi mx_1/L_x + 2\pi nx_2/L_y\right)},
\end{equation}

\noindent where \(\eta_{mn}(t) \in \mathbb{C}\) are the complex-valued Fourier series coefficients of the elevation at time \(t\). The elevation field is real-valued; therefore, the Fourier coefficients obey a complex conjugate symmetry condition, \(\eta_{m,n} = \eta_{-m,-n}^\ast\). The sum in \eqn{eq:orgf4173d0} is over an infinite set of frequencies. To make the synthesis computationally tractable, the series is truncated at \(M\) terms in the x direction and \(N\) terms in the y direction, leaving

\begin{equation}
\label{eq:orgd0ddea6}
\eta(\mathbf{x},t) = \sum_{m = -M/2}^{M/2-1}\sum_{n = -N/2}^{N/2-1} \eta_{m,n}(t) e^{i\left(2\pi mx_1/L_x + 2\pi nx_2/L_y\right)}.
\end{equation}

Using \eqn{eq:orgd0ddea6}, the elevation can be evaluated at any point within the patch with a cost of \(\mathcal{O}(MN)\). The Fourier series can be evaluated more efficiently on an \(M \times N\) rectangular grid using the fast Fourier transform (FFT). The FFT evaluates \eqn{eq:orgd0ddea6} at the points \(\mathbf{x}_{jk} = [L_x j/M,L_y k/N]\) for \(j \in [0,1,\dots,M-1]\) and \(k \in [0,1,\dots,N-1]\). FFT libraries such as FFTW \cite{Frigo_2005} refer to \eqn{eq:orgd0ddea6} as the ``inverse fast Fourier transform,'' which is computed in Julia or MATLAB as \texttt{ifft(eta)}, where the \(M \times N\) complex-valued array, \texttt{eta}, stores the Fourier coefficients.

The Karhunen-Lo\`{e}ve theorem \cite{Wong_1985} represents a random field as a randomly weighted sum of basis functions. In other words, a random field model for the seafloor elevation can be constructed using \eqn{eq:orgd0ddea6} with random Fourier coefficients \(\eta_{m,n}\). If the Fourier coefficients are modeled as independent, complex-normally distributed random variables with a mean of zero and a variance equal to the value of the power spectrum at the corresponding Fourier frequency, then the random field synthesized by \eqn{eq:orgd0ddea6} will be a stationary Gaussian random field with a covariance function given by the Fourier transform of the power spectrum. The procedure used by the seafloor synthesis module of NSEA to synthesize gridded seafloor representations is summarized as: %in Algorithm (\ref{org5b8ac38}).

\begin{algorithm}[h!]
\KwIn{$S_{mn}$: Power spectrum defined at the Fourier frequencies}
\KwOut{$\eta_{jk}$ : A random seafloor realization on the equally spaced $M \times N$ grid}
$\zeta_{mn} \sim CN(0,1)$\;
$\eta_{jk} = \sum_{m=-M/2}^{M/2-1}\sum_{n = -N/2}^{N/2-1} \sqrt{S_{mn}}\zeta_{mn} e^{i2\pi   m j/M+ 2\pi k/N}$\;
\caption{\label{org5b8ac38}
Seafloor synthesis algorithm}
\end{algorithm}

The final line is implemented in practice using a fast Fourier transform. Details on the the implementation of the fast Fourier transform and the scaling and arrangement of its coefficients are discussed in Section \ref{sec:org14563c3}.

This algorithm is efficient because of its use of the Fourier transform; however, there are several drawbacks to this method. First, the generated random fields are periodic, with periods \(L_x\) and \(L_y\) in the x and y directions. This periodicity induces spurious correlations between pixels on opposite sides of the patch. The simplest way to mitigate spurious correlations is to generate realizations over a patch at least twice as large as that needed to contain all of the observations \cite{Paciorek_2007}. For example, the roughness observations described further in Section \ref{sec:org2def406} are acquired with rotary scanning sonars with a range of 6 meters, so a patch with a side length of 12 meters centered on the instrument location contains all of the observations. However, setting \(L_x\) and \(L_y\) to 12 meters will result in spurious correlations at the edges of the patch, so the patch side lengths should be set to at least 24 meters. There are still spurious correlations between the edges of the simulated grid, but the correlations will not affect the region of the grid covered by the observations. Simulating on a larger grid is similar to the circulant embedding method \cite{Chan_1999,Dietrich_1997} used to simulate Gaussian processes on grids from a given covariance function. However, the circulant embedding method is more complex than the spectral synthesis algorithm, largely because it must modify the covariance function to ensure that it is nonnegative definite \cite{Stein_2002,Gneiting_2006}. A nonnegative definite covariance function corresponds to a power spectrum that is always nonnegative. Because the power spectrum is modeled directly, it is always nonnegative, and the additional complexity of circulant embedding techniques is not necessary.

The second drawback to the Fourier synthesis approach arises when observations are available on non-Cartesian grids. Rotary scanning sonars, for example, provide data on a polar grid defined by constant spacing in range and azimuth. Equation (\ref{eq:orgd0ddea6}) could be evaluated directly on the polar grid, but the computational efficiency of the fast Fourier transform is lost. To handle non-uniform observations, NSEA generates seafloor realizations on a uniform grid and then interpolates below the grid scale. Interpolation smoothes the elevation field at subgrid scales, so that the power spectrum of the interpolated field will not be identical to that of the true field. This discrepancy diminishes as the resolution of the grid is increased. As long as the resolved grid meets the Nyquist criterion of the dominant ripple field, inferences about the roughness should not be strongly affected by smoothing at high frequencies. Three interpolation schemes are used within NSEA (Fig. \ref{fig:org61d2c4c}). The selection of which scheme to employ depends on the model application. Nearest-neighbor interpolation \cite{Paciorek_2007} represents the seafloor with flat pixels defined by the grid (Fig. \ref{fig:org61d2c4c}a). This method is computationally efficient, but the resulting model of the seafloor is not continuous. Bilinear interpolation linearly interpolates along the x-axis and then along the y-axis, resulting in a continuous elevation model (Fig. \ref{fig:org61d2c4c}b). It is more computationally intensive than nearest-neighbor interpolation. However, bilinear interpolation is implemented natively in the texture filtering hardware of graphics processing units, which makes it competitive with nearest-neighbor interpolation at the cost of requiring more complex code to control the GPU. While the elevation is continuous using bilinear interpolation, the spatial derivative of elevation (i.e., seafloor slope) is not. Continuous spatial derivatives are especially useful when working with acoustic scattering models that depend on the local surface slope. Bicubic interpolation can be used in this case, providing continuous first and second derivatives of the elevation field (Fig. \ref{fig:org61d2c4c}c). Bicubic interpolation can also be implemented efficiently using texture filtering functions on the GPU \cite{Sigg_2006}. For qualitative comparisons between seafloor realizations and non-Cartesian data, nearest-neighbor interpolation suffices and can be computed efficiently on the CPU. For further modeling using the seafloor realizations, bicubic interpolation is recommended because of its continuous derivatives, but requires a GPU for efficient computation. An alternative to bicubic interpolation is to compute the desired derivatives of the seafloor elevation field using spectral methods \cite{Boyd_2001} and to interpolate the derivatives using nearest neighbors. This method can be very efficient, but does not result in a consistent model of seafloor elevation and may cause spurious artifacts when used in downstream modeling efforts.

\href{figures/interpolation\_figure.png}{\begin{figure}
\centering
\includegraphics[width=.9\linewidth]{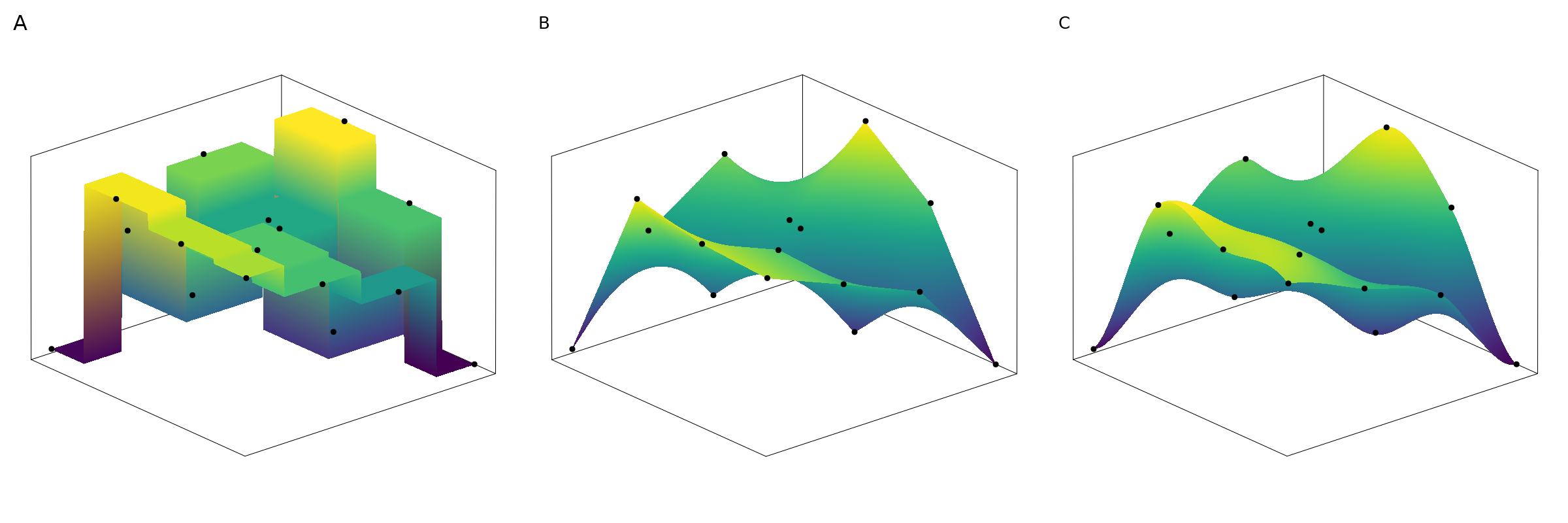}
\caption{\label{fig:org61d2c4c}Subgrid-scale interpolation schemes used within NSEA. The black dots represent the original gridded elevation measurements and are identical in each of the three plots. A. Nearest-neighbor interpolation, B. bilinear interpolation, C. bicubic interpolation}
\end{figure}}

\section{Observations}
\label{sec:orgee79e02}
The seafloor elevation field is never observed directly. It is only ever observed through indirect and noisy measurements. Extracting useful information about seafloor roughness and its evolution from these noisy and indirect observations requires a probabilistic observation model that connects the available observations to the seafloor elevation field. Statistical inference uses observations, the observation model, and the seafloor evolution model to estimate the seafloor elevation, statistical properties of the elevation field (e.g., power spectrum), or model parameters from the observations. 

To enable statistical inference of seafloor properties from observations, the observation model is formally specified as a conditional probability distribution \(\mathbf{z} \sim p_{\theta}(\mathbf{z}|\eta)\) for the observations, \(\mathbf{z} = [\mathbf{z}_1,\mathbf{z}_2,\dots,\mathbf{z}_N]\), given the realization of the seafloor, \(\eta\). The distribution is parameterized by a collection of parameters \(\theta\) that represent characteristics of the measurement system, such as the signal-to-noise ratio and the point spread function, or geophysical parameters, such as the sediment grain size. 

While the observation model is a conditional probability distribution, it is not always possible to evaluate the corresponding density function, \(p_{\theta}(\mathbf{z}|\eta)\), specifically when the observations depend on additional \emph{latent}, or unobserved, variables, \(\mathbf{u}\), in addition to the seafloor elevation. The observation probability density function is then determined by integrating the joint distribution of the observed and latent variables over the latent variables:

\begin{equation}
p_{\theta}(\mathbf{z}|\eta) = \int p_{\theta}(\mathbf{z},\mathbf{u}|\eta)\mathrm{d}\mathbf{u}.
\end{equation}

With the exception of a few special cases, this integral cannot be evaluated analytically, and numerical approximations must be made. However, random samples from the observation distribution can often be drawn easily from this model by first sampling the latent variables, \(\mathbf{u}\), from the probability distribution, \(p_{\theta}(\mathbf{u}|\eta)\), and then sampling the observed variables from the distribution \(p_{\theta}(\mathbf{z}|\mathbf{u},\eta)\). Such a \emph{generative} model enables statistical inference in complex models that do not have tractable likelihoods using specialized inference techniques \cite{Cranmer_2020}. Complex multiphysics simulations can therefore be treated within the same statistical framework as simpler, analytically tractable statistical models.

NSEA separates the observation model from the seafloor evolution model to enable many different kinds of observations to be used in estimating properties of the seafloor roughness. As long as the observation model can be represented in the form of a conditional probability distribution described above, statistical inference proceeds identically. Users who work with particular observation types can use NSEA to develop and test their own observation models with realistic seafloor roughness information provided by NSEA's seafloor evolution and synthesis models. 

Two observation models built into the reference implementation of NSEA are described in \secand{sec:org2873a64}{sec:org854c271}. The first model is a simplified observation model appropriate for use with observations from profiling sonars, stereophotography, structure-from-motion, or other measurement systems that return high-resolution bathymetric elevation information. This simplified model illustrates the considerations that go into designing an observation model. The second is an imaging sonar model used to compare acoustic return data from rotary scanning sonars to the output of the seafloor evolution model. The imaging sonar model is not an accurate seafloor acoustic model, but it does enable qualitative comparisons between observations and model outputs, and it demonstrates how more sophisticated acoustic models could be incorporated into NSEA.

\subsection{A Simplified Observation Model}
\label{sec:org2873a64}
Profiling sonars are often used to make high-resolution elevation maps in sediment transport studies \cite{Traykovski_2007,Brakenhoff_2020}. These sonars have very narrow beams in both azimuth and elevation, so the ensonified patch of seafloor is relatively small, especially at short ranges. The measurements recorded by the profiling sonar are a time series of backscattered signal strength for each sonar pulse at a given azimuth and elevation. The time axis records the two-way travel time of the transmitted pulse from the sonar to the seafloor, where it is then reflected back to the sonar. The profiling sonar identifies the seafloor by selecting the time of the first strong return in the time series. Knowing the speed of sound in water, \(c\), the two-way travel time, \(t\), can be converted to range from the sonar, \(R\), by the formula \(R = ct/2\). The processed data from the profiling sonar then consists of a set of azimuth \((\theta)\) and elevation \((\varphi)\) angles and the range of the first return from the sonar, \((\theta_i,\varphi_j,R_{ij})\). Standard trigonometry converts these polar coordinates into Cartesian coordinates, producing a collection of horizontal points and corresponding seafloor elevation, \((\mathbf{x}_{ij},\eta_{ij})\). 

\begin{equation}
z_i = \int h(\mathbf{x}_i,\mathbf{x}^\prime) \eta(\mathbf{x}^\prime)\ \mathrm{d}\mathbf{x}^\prime = (h \ast \eta).
\end{equation}

A true point observation of the field has a point spread function equal to a Dirac delta function \(h(\mathbf{x},\mathbf{x}^\prime) = \delta(\mathbf{x} - \mathbf{x}^\prime)\). A more realistic point spread function with a finite width will effectively blur the seafloor elevation field, smoothing out high-frequency variability. This indirect measurement model may not be significant when working with a narrow-beamed profiling sonar close to the seabed, where the point spread function will be narrow enough not to smooth away important roughness information, but in other observational scenarios or modalities, it might not be possible to resolve small-scale roughness because of the inherent properties of the measuring device.

The point spread function represents the ``known'' measurement process in the sense that it is intrinsic to the instrument and is reproducible and deterministic. However, even with perfect knowledge of the instrument's point spread function, other processes will contaminate the measurements. For example, the first return may represent scattering from organisms in the water column rather than from the seafloor, or the speed of sound might fluctuate with changes in water temperature. The data will always have ``noise'' in that the observation, \(z_{i}\), acquired from the sonar at a point \(\mathbf{x}_{i}\) is equal to the seafloor elevation plus some residual error:
\begin{equation}
\label{eq:orgce68a92}
z_{i} = (h \ast \eta)(\mathbf{x}_i) + \varepsilon_{i}.
\end{equation}
The measurement error terms are unobserved by definition; however, if they were known, they could be corrected in the data-processing pipeline. A common model for the error term is an assumption of independent, normally distributed measurement errors with zero mean and variance, \(\sigma_\varepsilon^2\), in which case the conditional distribution of the observation, \(z_i\), given the seafloor elevation, \(\eta_i\), is also normal:
\begin{equation}
z_{i} | \eta \sim \mathcal{N}((h \ast \eta)(\mathbf{x}_i,\sigma_\varepsilon^2).
\end{equation}
The observation model parameters are the point spread function, \(h\), and the noise variance, \(\sigma_\varepsilon^2\). This model is not necessarily an accurate model of the measurement errors in profiling sonar observations, but it does allow for analytical statistical treatment. For example, since NSEA assumes \(\eta\) to be a Gaussian process with a known power spectrum, \(S(k)\), the Wiener filter \cite{Wiener_1949} optimally estimates the seafloor elevation from the observations. More complicated measurement models could be considered; for example, errors could depend on the location or could be correlated, or error distributions could have heavy tails that would accommodate outliers.

\subsection{Imaging Sonar Observation Models}
\label{sec:org854c271}
Imaging sonars are widely used due to their ability to observe large areas of the seafloor at a high resolution. However, their measurements are very different from those of the profiling sonars described above. Similar to profiling sonar, an imaging sonar ensonifies an area of the seafloor and records a time series of backscattered signal strength. It differs from profiling sonar in the shape of the ensonified area and in the processing of the time series to the final observations. Profiling sonars have narrow beams of a few degrees around a central bore, whereas imaging sonars have wide beams in the elevation direction. Side-scan and rotary scanning sonars have narrow beams in the azimuthal direction to ensonify narrow strips of seafloor. Synthetic aperture sonars use wide azimuthal beams, but process multiple pulses together to synthesize narrow beam widths. The instrument samples the backscattered signal from a single pulse at discrete times, so each sample corresponds to a particular two-way travel time from the instrument. This travel time is converted into range from the sonar using the same \(R = ct/2\) formula in processing profiling sonar measurements, and the output of the instrument is essentially a series of backscattered signal amplitudes at a set of discretely spaced ranges for each pulse. The instrument then moves through the water, so the next pulse ensonifies a different portion of the seafloor. Side-scan sonars move forward, mapping out strips on either side of the instrument, while rotary-scanning sonars rotate in the azimuthal direction and map an annulus around the instrument position. A complete scan consists of a stack of backscattered amplitude time series. These stacks can then be projected into geographic coordinates using knowledge of the instrument's location to produce a sonar image.

Sonar images are not maps of seafloor elevation, yet bathymetric features like ripples in sonar imagery can be observed because the backscattered amplitude depends on the orientation of the seafloor surface relative to the instrument. The simplest model for the angular dependence of the scattering strength is given by Lambert's law for diffuse scattering, which roughly states that the acoustic energy impinging on the seafloor is scattered equally in all directions \cite{Bell_1999,Jackson_2007}. Taking into account the relative orientation of the seafloor surface to the instrument gives the well-known cosine-squared form of Lambert's law for the backscattering cross section, \(\sigma\),
\begin{equation}
\label{eq:org7365d05}
\sigma = \frac{\rho}{\pi} \cos^{2} \theta_i,
\end{equation}
\noindent where \(\rho\) is the proportion of incident energy that is scattered rather than absorbed by the seafloor, and \(\theta_i\) is the incident angle relative to the surface normal. This model is often given in the acoustic literature with the grazing angle, \(\theta_g\), relative to the seafloor surface, in which case \(\sigma \propto \sin^2 \theta_g\).

The cosine of the incident angle can be written as \(\cos \theta_i = \hat{\mathbf{n}} \cdot \hat{\mathbf{L}}\), where \(\hat{\mathbf{n}}\) is the unit surface normal vector and \(\hat{\mathbf{L}}\) is the unit vector pointing from the surface to the sonar. If the seafloor surface is given as \(\eta(\mathbf{x})\) and the sonar location is \(\mathbf{o}\), then the surface normal vector is given by
\begin{equation}
\label{eq:org4ac21a2}
\mathbf{n} = \left[ -\frac{\partial \eta}{\partial x}, -\frac{\partial \eta}{\partial y}, 1\right]
\end{equation}
\noindent and the observation direction vector is \(\mathbf{L} = \left[o_x - x, o_y - y, o_z - \eta(x,y) \right]\). The unit vectors are obtained by normalizing these vectors.

Lambert's law can form the basis of a simplified imaging sonar observation model within NSEA, as it can simulate the backscattering cross section that would be obtained from a given seafloor roughness realization. For rotary scanning sonars, the natural range and azimuth coordinates of the acquired data can be converted into Cartesian surface coordinates with trigonometry, assuming that the amplitude of seafloor roughness is small relative to the height of the sensor above the seafloor. The required seafloor elevation field and its derivatives can be computed efficiently with the fast Fourier transform and filtering procedures outlined in Section \ref{sec:orge5b6faa}. Two further points must be considered to turn the Lambertian scattering model into a valid observation model for use in NSEA. First, instrument-specific effects such as the sonar transmit and receive beam patterns create additional variability in the image not due to the seafloor roughness \cite{Cervenka_1993,Bell_1997b}. These effects, however, affect the image only in the downrange, or across-track, direction, so they can be modeled as a range-dependent amplitude, \(B(r)\), which is multiplied by the backscattering cross section corresponding to the given range and azimuth angle. The image is then formed by the equation,
\begin{equation}
\label{eq:org808ac59}
Y(r,\theta) = B(r)\sigma(r,\theta),
\end{equation}
\noindent where \(\sigma(r,\theta)\) is the scattering cross section of the portion of the seafloor surface that corresponds to the given range, \(r\), and azimuth, \(\theta\).

The imaging sonar model finally needs to be written probabilistically to fit into NSEA's statistical inference framework. There are many models for the noise statistics of acoustic backscattering from the seafloor \cite{Lyons_1999,Abraham_2002,Jackson_2007,Lyons_2009}, but the simplest can be derived by assuming that the received sonar signal consists of contributions from a very large number of small scatterers. The central limit theorem suggests that the received signal will be approximately complex normally distributed, so the signal amplitude is Rayleigh distributed \cite{Lyons_1999}, and the probabilistic observation model can be written as \(Y(r,\theta) \sim \text{Rayleigh}(B(r)\sigma(r,\theta))\). The parameters of this observation model are contained entirely within the range-dependent amplitude function, \(B(r)\), which must be estimated from the data.

The Lambertian model is not necessarily a model of seafloor acoustics \cite{Bell_1999,Folkesson_2020}, and inferences on seafloor roughness from this observation model should be subjected to extensive testing and validation. However, it does enable qualitative comparisons between sonar imagery and the output of the seafloor evolution model that account for the characteristics of imaging sonar observations, namely the directional filtering of seafloor roughness \cite{Bell_1999}.

\chapter{Statistical inference}
\label{sec:orga29e677}
The previous section has described NSEA as a set of coupled models: forcing, seafloor evolution, seafloor synthesis, and observations. Each of these models can be described probabilistically, defining a probability distribution over its output conditional on its input. NSEA therefore takes the form of a hierarchical model:
\begin{align}
\label{eq:orgd926d1a}
\text{parameters} &\sim p_{\text{parameters}}(\cdot) \\
\text{forcing} &\sim p_{\text{forcing}}(\cdot | \text{parameters}) \\
\text{spectrum} &\sim p_{\text{NSEM}}(\cdot | \text{parameters}, \text{forcing}) \\
\text{seafloor} &\sim p_{\text{synthesis}}(\cdot | \text{spectrum},\text{parameters}) \\
\label{eq:observationlikelihood}
\text{observations} &\sim p_{\text{obs}}(\cdot | \text{seafloor}, \text{forcing}, \text{parameters}).
\end{align}
A few examples of inferences one might make using NSEA are:
\begin{itemize}
\item estimation of properties of the seafloor (e.g., grain size) from sidescan sonar imagery,
\item comparison of multiple formulations of sediment transport processes within the seafloor evolution model to assess their validity, and
\item forecast of the evolution of seafloor roughness at a particular location using forcing conditions predicted by a wave model.
\end{itemize}

These examples can be categorized by three types of statistical inferences: estimation, model comparison, and prediction, respectively. Their commonalities include the use of data to estimate unobserved quantities of interest and the quantification of the uncertainty in those estimates. Bayesian statistics provide a framework for making these inferences in hierarchical models. The goal of Bayesian inference is to use the observed variables to compute the \emph{posterior distribution} of the unobserved or latent variables. The posterior distribution can be obtained formally with Bayes's rule, which states that the posterior is proportional to the \emph{likelihood} (the observation model) times the \emph{prior} (the model for the latent variables). For example, the posterior distribution for the seafloor elevation field could be written as
\begin{equation}
p(\text{seafloor}|\text{observations}) = \frac{p(\text{observations}|\text{seafloor})p(\text{seafloor})}{p(\text{observations})},
\end{equation}
\noindent where the likelihood is \(p(\text{observations}|\text{seafloor})\) and the prior is \(p(\text{seafloor})\). Note, however, that the likelihood, \(p(\text{observations}|\text{seafloor})\), the prior, \(p(\text{seafloor})\), and the normalizing constant \(p(\text{observations})\) (also called the \emph{marginal likelihood}) do not appear directly in the hierarchical model, \eqn{eq:orgd926d1a}. They must be obtained by integrating or \emph{marginalizing} over the other variables in the model: in this case, the parameters, forcing and seafloor spectrum. For example, the prior distribution of the seafloor elevation is formally given by
\begin{equation}
\label{eq:orgae17c0b}
\begin{aligned}
p(\text{seafloor}) = \int & p(\text{seafloor}|\text{spectrum},\text{parameters})p(\text{spectrum}|\text{forcing},\text{parameters}) \\
& p(\text{forcing}|\text{parameters}) p(\text{parameters}) d\text{spectrum} d\text{forcing} d\text{parameters}.
\end{aligned}
\end{equation}
These integrals will not generally have analytical solutions, and they are very high-dimensional, so calculating them using numerical approaches like quadrature scale very poorly. Bayesian statistics offers a wide variety of algorithms for estimating integrals like \eqn{eq:orgae17c0b}. The choice of algorithm depends on the kinds of observations that are available, the posterior distributions that are of interest, and the simplifications that the user is willing make to the complete hierarchical model. One algorithm, approximate Bayesian computation \cite{Sunnaker_2013}, that is implemented in NSEA is described.

\section{Approximate Bayesian Computation in NSEA}
\label{sec:org5de6ca7}
The inference task considered in this section is to estimate the parameters of the seafloor evolution model from observations of ripple wavelength that have been extracted from sonar imagery. This section describes the methodology, and results from a test of this methodology are presented in Section \ref{sec:org2def406}. The Bayesian inference problem is to estimate the posterior distribution of the parameters given the wavelength observations
\begin{equation}
\label{eq:orgfe91c59}
p(\theta | \hat{\lambda}_{1:T}, u_{1:T}, A_{1:T}) = \frac{p(\hat{\lambda}_{1:T}|\theta, u_{1:T}, A_{1:T}) p(\theta)}{p(\boldsymbol{\lambda}_{1:T})},
\end{equation}
\noindent where the parameters are collected in the vector, \(\theta\), and the observations consist of a time series of estimated ripple wavelengths, \(\hat{\lambda}_{1:T}\). The case presented in Section \ref{sec:orga29e677} uses a fast Fourier transform to process the sonar images and to estimate the peak ripple wavelength \cite{Penko_2017}, but other image processing techniques \cite{Traykovski_2007,Skarke_2011} could also be employed. The forcing information is supplied as a time series of bottom orbital velocity, \(u_{1:T}\), and semiorbital excursion, \(A_{1:T}\). The prior distribution for the parameters, \(p(\theta)\), is ultimately chosen by the user based on assumptions about the ranges of the parameters. 

The likelihood, \(p(\hat{\lambda}_{1:T}|\theta, u_{1:T}, A_{1:T})\), is the probability distribution over the wavelength data induced by the hierarchical model. However, this likelihood cannot be analytically computed from \eqns{eq:orgd926d1a}{eq:observationlikelihood} for two reasons. First, it requires a high-dimensional integral over the seafloor realization that cannot be computed analytically. It is possible to estimate this integral using traditional Bayesian inference algorithms like Markov chain Monte Carlo. The more challenging problem is that the morphological processing required to estimate ripple wavelengths from the sonar imagery applies a complicated, nonlinear, and non-invertible function to the images. This additional processing step does not allow for the computation of the probability distribution for the estimated ripple wavelengths conditioned on the seafloor realization, \(p(\hat{\lambda} | \eta)\), which is needed for Monte Carlo sampling of the seafloor realization.

Approximate Bayesian computation (ABC) \cite{Sunnaker_2013} enables Bayesian inference for models with intractable likelihoods by replacing direct computation of the likelihood with a simulation. In the context of the parameter inference problem given in \eqn{eq:orgfe91c59}, ABC consists of sampling parameter values from the prior distribution, \(p(\theta)\), running the model forward to simulate ripple wavelengths, and then comparing the simulated ripple wavelengths with the observed ripple wavelength by calculating a discrepancy measure such as the mean squared difference between the simulated and observed ripple wavelengths. Parameter values with a corresponding discrepancy measure less than a user-defined tolerance are accepted by the algorithm, and the posterior distribution is approximated by the accepted parameter values. Approximate Bayesian computation is exact in the limit as the discrepancy tolerance goes to zero, but a tolerance of zero rejects almost all of the sampled parameter values, which means more samples must be taken to reduce the statistical variance of the posterior estimate. Conversely, a tolerance set too high will accept all of the sampled parameters, and the resulting posterior will more closely resemble the prior distribution than the true posterior distribution. Choosing the tolerance requires making a trade-off between the bias of a too-high tolerance and the variance of a too-low tolerance. A pilot simulation with relatively few samples can help determine the acceptance rate as a function of the tolerance. The tolerance is then set to acquire the desired number of accepted samples given the total number of samples that can be run with the available computational budget.

To simulate ripple wavelengths, one would first run the seafloor evolution model with the sampled parameter value to estimate the seafloor spectrum, synthesize a seafloor realization from that spectrum, apply the sonar imagery model to simulate a sonar image consistent with the spectrum, and finally, use the image processing method of choice to estimate ripple wavelengths from the simulated sonar image. However, this complete simulation is computationally intensive and the inferences depend heavily on the accuracy of the sonar imagery model. NSEA currently simulates ripple wavelengths by calculating the peak ripple wavelength from the maximum of the simulated seafloor spectrum and adding noise to the peak wavelength value. Future work is needed to enhance the computational capacity of NSEA and to enhance the realism of the sonar imagery model to allow ABC to use complete simulations from the hierarchical model.

\chapter{Implementation}
\label{sec:orgc193e3b}
Development of NSEA currently focuses on an implementation written in the Julia language \cite{Bezanson_2017} called Stingray.jl. Stingray provides a core interface for describing the seafloor elevation within a given localized area using spectral methods (StingrayCore, Section \ref{sec:orgb0a1f96}), an implementation of the spectral seafloor evolution model (NSEM, Section \ref{sec:orgb0e600a}), a set of methods for synthesizing seafloor realizations from spectral information (SyntheticSeafloor, Section \ref{sec:org14563c3}), and an imaging sonar model for simulating sidescan sonar images from seafloor realizations (Stingray, Section \ref{sec:org82c9ad0}). Each of these components is implemented as a separate Julia package, but are developed within a single Git repository that facilitates installing compatible versions of each of the packages.
\section{StingrayCore}
\label{sec:orgb0a1f96}
The StingrayCore module provides a lightweight interface for representing seafloor elevation realizations within a rectangular patch. The module exports a single concrete type:
\clearpage
\begin{Verbatim}[frame=single]
struct Seafloor{T, D,
                A <: AbstractArray{T,D},
                B <: AbstractArray{Complex{T},D},
                P <: AbstractFFTs.Plan{T}}
    p::P
    F::B
    V::B
    eta::A
    kx
    ky
    dims
    L::Tuple{T,T}
end
\end{Verbatim}

The fields of this type are a Fourier transform plan, \texttt{p}, as defined in the AbstractFFTs.jl interface, fields for the Fourier transform of the elevation, \texttt{F}, and a filter impulse response, \texttt{V}, that is used to generate realizations having a given power spectrum. The field \texttt{eta} is a real-valued array that contains a gridded seafloor elevation. These three arrays, \texttt{F,V,eta}, are not necessary for running the seafloor evolution model, but preallocating them in the \texttt{Seafloor} prevents having to reallocate potentially very large arrays when synthesizing seafloor realizations. The geometry of the patch is defined by the fields \texttt{kx}, \texttt{ky}, \texttt{dims = (M, N)} and \texttt{L = (Lx,Ly)}, which are the wavenumber grids in the x and y directions, the dimensions of the gridded seafloor representation, and the length of the patch sizes in the x and y directions in meters. A \texttt{Seafloor} object is created by calling the type constructor \texttt{Seafloor(u,L)} with a prototype array, \texttt{u}, that has the same data type and size as the desired gridded seafloor representation, and the tuple \texttt{L = (Lx,Ly)} of patch side lengths. The prototype array \texttt{u} is only used to determine the correct sizes of the various internal fields of the \texttt{Seafloor} type. The use of the the prototype array allows diverse array types from the Julia ecosystem to be used as storage for the gridded seafloor representation and its Fourier coefficients. However, a user of Stingray most likely would use a standard Julia array with an element type of \texttt{Float64} and a size \texttt{(M,N)}, in which case the desired \texttt{Seafloor} object can be created as:

 \noindent \texttt{sf = Seafloor(zeros(M,N),(Lx,Ly))}. 
 
 \noindent Two convenience methods are also provided to allow users to create \texttt{Seafloor} objects without worrying about the details of prototype array: 
 
 \noindent \texttt{sf = Seafloor((M,N),(Lx,Ly))} 
 
\noindent  is equivalent to 
 
\noindent \texttt{sf = Seafloor(zeros(M,N),(Lx,Ly))} 
 
\noindent  and 
 
\noindent  \texttt{sf = Seafloor(N,L)} 
 
\noindent  is equivalent to 
 
\noindent  \texttt{sf = Seafloor(zeros(N,N),(L,L))}. 
 
\noindent  The patch and its grid representation have equal lengths and resolutions in the x  and y directions using the latter method.

The only restriction on the prototype array is the requirement for it to be an input to the \texttt{plan\_rfft(u)} to create a FFT plan. For the regular Julia \texttt{Array}, the FFTW.jl bindings to the FFTW library \cite{Frigo_2005} provide this plan. One must load the desired FFT provider before attempting to create a \texttt{Seafloor} object, or the \texttt{Seafloor} command will result in an error. See the below example:
\begin{Verbatim}[frame=single]
# Load in the required Julia packages
using StingrayCore

# Create the seafloor object
M,N=(256,512)
Lx,Ly=(5.0,10.0)

# Load the FFT provider
using FFTW

# Create the seafloor object with the method for rectangular patches
sf = Seafloor((M,N),(Lx,Ly))
\end{Verbatim}

\noindent or

\begin{Verbatim}[frame=single]
# Create the seafloor object with the method for square patches
sf = Seafloor(M,L)
\end{Verbatim}

The various models can also be run on CUDA-enabled GPUs by using the \texttt{CuArray} type from the CUDA.jl package \cite{Besard_2018}. One could create a \texttt{Seafloor} on the GPU with \texttt{sf\_gpu = Seafloor(CUDA.zeros(M,N), (Float32(Lx), Float32(Ly)))}. Note the casting of the patch sizes to \texttt{Float32}, which is necessary because the default element type for a \texttt{CuArray} is a 32-bit floating point number. If \texttt{Lx} and \texttt{Ly} are another type, such as \texttt{Float64}, type instabilities will result, which can dramatically slow down operations on the GPU. The type parameters of the \texttt{Seafloor} enforce these types to match. If one tries to pass a prototype array with \texttt{Float32} elements and \texttt{Float64} patch edge lengths, a method error will result. The \texttt{Seafloor((M,N),(Lx,Ly))} convenience methods automatically create a Julia \texttt{Array}-based \texttt{Seafloor} object, so the method with the prototype array must be used to create CUDA-enabled \texttt{Seafloor} objects.

For CuArrays, CUDA.jl provides an interface to FFTs through CUFFT. To employ CUDA-enabled GPUs, the following commands can be used:

\begin{Verbatim}[frame=single]
using CUDA
M,N=(256,512)
Lx,Ly=(5f0,10f0)
sf_gpu = Seafloor(CUDA.zeros(M,N),(Lx,Ly))
\end{Verbatim}

\section{NSEM}
\label{sec:orgb0e600a}
The NSEM module implements the spectral seafloor evolution model described in Section \ref{sec:org5ad68e6}. The module exports a single function \texttt{nsem}, which has multiple methods used to run the model with different configurations. The most general method is \texttt{nsem(sf::Seafloor, t, uw, Aw, phiw, uc, phic, params)}, which takes a \texttt{Seafloor} object, \texttt{sf}, a vector of time steps, \texttt{t}, given in seconds since an arbitrary time, and similarly sized vectors of bottom wave orbital velocity (\texttt{uw} $\left[ m/s \right]$), semiorbital excursion (\texttt{Aw} $\left[ m \right]$), wave direction (\texttt{phiw} $\left[\textrm{radians counterclockwise from the x-axis}\right]$), bottom current velocity (\texttt{uc} $\left[ m/s \right]$), and bottom current direction (\texttt{phic} $\left[ \textrm{radians counterclockwise from the x-axis}\right]$). The model parameters, \texttt{params}, are passed to the model in a \texttt{NamedTuple} with the following fields:
\begin{verbatim}
params = (;
    D50, # Median sediment grain size
    S, # Sediment specific gravity
    theta_cr,# Critical Shields stress for initiation of motion
    wo_diff, # logarithm of the difference between the 
              # critical stress for washout and for initiation 
              # of motion (e.g., log(0.168-0.05))
    alpha, # Time scale multiplier
    phi, # Porosity
    sigma, # Equilibrium spectral width
    gamma, # Sediment transport rate multiplier
    xi, # Sediment transport rate exponent
    D, # Diffusion coefficient
    Cd, # Drag coefficient
    rp # Ripple predictor
)
\end{verbatim}

Most of these parameters are simply real numbers, though care should be taken to ensure that they have the same type as the underlying array type of the \texttt{Seafloor} type (e.g., \texttt{Float64} if an \texttt{Array\{Float64\}} is used, \texttt{Float32} if a \texttt{CuArray\{Float32\}} is used). The washout threshold must always be greater than the critical Shields stress for initiation of motion, \texttt{theta\_cr}, so the parameter input for the washout threshold is the logarithm of its excess over the critical Shields stress (\texttt{wo\_diff}). The washout threshold, \texttt{theta\_wo}, is computed within NSEM as:

\noindent \texttt{theta\_wo = params.theta\_cr + exp(params.wo\_diff)}. 

\noindent The ripple predictor is an object that is a subtype of \texttt{NSEM.EquilibriumRipplePredictor}. Each \texttt{EquilibriumRipplePredictor} object is callable with a method,

%\begin{Verbatim}[frame=single]
\noindent \texttt{(rp::EquilibriumRipplePredictor)(Ab,psi,theta,D50)},
%\end{Verbatim}

\noindent that takes the semiorbital excursion, \texttt{Ab}, the mobility number, \texttt{psi}, the Shields stress, \texttt{theta}, and the grain size, \texttt{D50}, and returns \texttt{(eta, lambda)} the equilibrium ripple height and wavelength. Five equilibrium ripple predictors are currently implemented: \texttt{Nielsen} \cite{Nielsen_1981}, \texttt{WibergHarris} \cite{Wiberg_1994}, \texttt{Goldstein} \cite{Goldstein_2013}, \texttt{Nelson} \cite{Nelson_2013}, and \texttt{ParameterizedRipplePredictor}. The first four of these are empirical formulas from the literature. The relevant ripple predictor objects can be created as \texttt{Nielsen()}, \texttt{WibergHarris()}, etc., and passed as a model parameter in \texttt{params}, to \texttt{nsem}. The parameterized ripple predictor represents the power law model of Eqs. \eqref{eq:org13961b3} and \eqref{eq:org13961b3a}. A \texttt{ParameterizedRipplePredictor} object is created by passing vectors with the exponents for each of these terms, \texttt{beta} and \texttt{chi}. 

The current-generated ripple predictor from Soulsby and Whitehouse \cite{Soulsby_2012} has also been implemented (\texttt{SoulsbyWhitehouse}). It requires the sediment specific gravity, \texttt{S}, which is passed to the object constructor as \texttt{rp = SoulsbyWhitehouse(S)}. Under current-dominated conditions, the semiorbital excursion is undefined, and the method for the Soulsby and Whitehouse ripple predictor only takes three parameters, 

\noindent \texttt{(erp::SoulsbyWhitehouseCurrent)(psi,theta,D50)}. 

\noindent The Soulsby and Whitehouse predictor depends only on the grain size and the sediment's specific gravity, but the forcing variables \texttt{psi} and \texttt{theta} are retained to provide compatibility with future ripple predictor implementations that may depend on the strength of the current forcing. When the shear stress due to currents is stronger than that due to waves, the \texttt{SoulsbyWhitehouse} ripple predictor is used to compute the equilibrium ripple height and wavelength. The parameter \texttt{params.rp} should always be one of the wave-generated ripple predictors, even for current-dominated systems. The \texttt{SoulsbyWhitehouse} predictor will automatically be used for current-generated ripples. Future versions of NSEM might allow the specification of alternative current-generated ripple predictors.

See \Tbl{tab.params_typ} for a list of typical values and units of the model parameters.

\begin{table}[htbp]
\caption[Typical NSEM model parameter values and units]{Typical NSEM Model Parameter Values and Units}
\label{tab.params_typ}
\centering
\begin{tabular}{|c|c|c|}
\hline
Parameter & Units & Typical Value\\
\hline
\hline
\(D_{50}\) & m &  0.00023 \\
S & - & 2.65\\
\(\theta_{cr}\) & - &  0.05\\
\(\Delta\theta_{wo}\) & - & $\textrm{log}(\theta_{wo}-\theta_{cr})$\\
\(\alpha\) & - &10.0\\
\(\varphi\) & - & 0.4\\
\(\sigma\) & m & 1.0\\
\(\gamma\) & - & 8.0\\
\(\xi\) & - & 1.5\\
\(D\) & m\(^{\text{2}}\) s\(^{\text{-1}}\) & $5.0\times10^{-9}$ \\
Ripple predictor & - & Nelson() \\ 
Grid size &  - & 512 \texttimes{} 512\\
Patch edge length & m & 15 \\
\hline
\end{tabular}
\end{table}

\noindent The output of:

\noindent \texttt{nsem( sf::Seafloor, t, uw, Aw, phiw, uc, phic, params )} 

\noindent is a three-dimensional array containing the simulated amplitude spectrum. The first two dimensions are the wavenumber in the x and y directions and have lengths equal to the lengths of \texttt{sf.kx} and \texttt{sf.ky}, respectively. The third dimension represents the time step and has a length equal to the length of the vector \texttt{t}, which should be equal to that of \texttt{uw}, \texttt{Aw}, etc.

\noindent An \texttt{nsem} method for running NSEM with only wave-generated ripples is also provided:

\noindent \texttt{nsem( sf::Seafloor, t, uw, Aw, phiw, params )}. 

\noindent If the system of interest is wave-dominated or current data are not available, one can provide \texttt{nsem} with only wave forcing data. The parameters, \texttt{params}, are the same as for the full model, except that the drag coefficient, \texttt{Cd}, is not required and will be ignored if provided. To ease integration between NSEM and other sources of forcing information, a method, 

\noindent \texttt{nsem( sf::Seafloor, df, params )}, 

\noindent is provided for running the model from forcing data within a \texttt{DataFrame} by reading a CSV file containing the forcing information. The columns of the DataFrame should be labeled as ``\texttt{t}'', ``\texttt{uw}'', ``\texttt{Aw}'', ``\texttt{phiw}'', ``\texttt{uc}'', and ``\texttt{phic}'', corresponding to the inputs passed to the complete method of \texttt{nsem}. The ``\texttt{t}'' column should consist of \texttt{DateTime} objects from the Julia standard library Dates module that record the observation times.

\section{SyntheticSeafloor}
\label{sec:org14563c3}
The SyntheticSeafloor module implements Fourier synthesis of seafloor elevation realizations consistent with a given amplitude spectrum. The Fourier synthesis algorithm is discussed in Section \ref{sec:orge5b6faa}. This section discusses the implementation of this algorithm and its use with other Stingray packages. SyntheticSeafloor exports two functions, 

\noindent \texttt{synthesize( sf::Seafloor, v, u )}

\noindent and 

\noindent \texttt{synthesize!( sf::Seafloor, v, u )}. 

\noindent Both functions expect a \texttt{Seafloor} object (discussed in Section \ref{sec:orgb0a1f96}) that defines the geometry of the grid (an \(M \times N\) array) and two arrays, \texttt{v} and \texttt{u}. Array \texttt{v} is the desired amplitude spectrum, organized in the pattern expected by FFTW for real-valued Fourier transforms. The output of NSEM, for example, is an array with dimensions \((\lfloor M/2 \rfloor +1,N,T)\) containing the amplitude spectrum at each time step of the simulation. To generate a seafloor elevation realization at time index \texttt{t} of the simulation, one would pass the slice \texttt{v[:,:,t]} to \texttt{synthesize}. Array \texttt{u} is an \(M \times N\) array of independent standard normal variables, generated, for example, by \texttt{randn(M,N)}. Both \texttt{synthesize} and \texttt{synthesize!} then compute the gridded realization of the seafloor elevation, \texttt{eta} as:

%\begin{Verbatim}[frame=single]
\noindent \texttt{eta = irfft(v .* rfft(u),sf.dims[1])}.
%\end{Verbatim}

The difference between \texttt{synthesize} and \texttt{synthesize!} is in how this computation is performed internally. \texttt{synthesize} implements the seafloor synthesis algorithm using the Fourier transform plan cached in the \texttt{sf} object

\begin{Verbatim}[frame=single]
function synthesize(sf::Seafloor,v,u)
    p = sf.p
    p \ (v .* (p * u))
end.
\end{Verbatim}

This implementation has the downside of allocating memory for the forward Fourier transform of \texttt{u}, the filter Fourier coefficients, and the gridded seafloor realization. Allocating memory repeatedly inside iterative algorithms such as Markov chains Monte Carlo inference can harm performance; therefore, the \texttt{synthesize!} variant is also provided. Following Julia conventions that functions ending in \texttt{!} modify their arguments, \texttt{synthesize!} performs in-place Fourier transforms using cached arrays that are created when the \texttt{Seafloor} object is created. 

\begin{Verbatim}[frame=single]
function synthesize!(sf::Seafloor,v,u)
    p = sf.p
    F,eta = sf.F,sf.eta

    # Perform the in-place forward FFT of u
    mul!(F,p,u)

    # Multiply by the amplitude spectrum in place
    F .*= v

    # Perform the in-place inverse FFT of F
    ldiv!(eta,p,F)

    eta
end
\end{Verbatim}

This method allocates no memory and is suitable for use in high-performance inner loops. However, because it modifies the arrays stored in the \texttt{Seafloor} object, it is not thread-safe: if two threads attempt to use \texttt{synthesize!} simultaneously on the same \texttt{Seafloor} object, the results will not be correct. If parallel processing is desired, either the non-mutating \texttt{synthesize} can be used or multiple identical \texttt{Seafloor} objects can be created and passed to each processor.

Both versions work for any fast Fourier transform that implements the \texttt{AbstractFFTs} interface. CUDA-accelerated seafloor synthesis requires constructing a \texttt{Seafloor} object on the GPU as discussed in Section \ref{sec:orgb0a1f96} and then passing \texttt{CuArrays} for \texttt{v} and \texttt{u}. The arrays can either be CPU arrays that are copied to the GPU using the \texttt{cu} function provided by \texttt{CUDA} (e.g., \texttt{synthesize(sf,cu(v),cu(u))} ), or they can be native GPU arrays. Random values for \texttt{u} can be generated on the GPU using \texttt{CUDA.randn(M,N)}, while if NSEM is run with a GPU-based \texttt{Seafloor} object, the output amplitude spectra will already be on the GPU and can be accessed using the same slice syntax as the CPU version. Other array types should work similarly, but are not regularly tested during Stingray development.

\section{Demo Code}
\label{sec:demo}

    The following example commands can be used to run the demo code for NSEM (\Secn{sec:orgb0e600a}) and the resulting time series of seafloor elevation realizations generated with the SyntheticSeafloor (\Secn{sec:org14563c3}). The demo code uses a time series of wave (only) forcing to run NSEM and then uses SyntheticSeafloor with the random noise, \texttt{u}, correlated in time with a correlation coefficient, \texttt{rho}. \\

\begin{lstlisting}[breaklines=true,frame=single]
using SyntheticSeafloor, NSEM, FFTW, DataFrames, CSV, DelimitedFiles

# Create Seafloor object
sf=Seafloor((512,512),(5.0,5.0));

# Load in wave forcing data as a DataFrame
df=DataFrame(CSV.File("NSEM_DEMO_DATA.csv"));
  
#  Set model parameters in a Named Tuple, $\uptheta$
$\boldmath{\uptheta}$=(D50 = 2.3e-4, S = 2.65, $\boldmath{\uptheta}$c = 0.05, $\boldmath{\upDelta \uptheta}$wo = log(0.168-0.05), $\boldmath{\upalpha}$ = 10.0, 
  $\boldmath{\upphi}$ = 0.4, $\boldmath{\upsigma}$ = 1.0, $\boldmath{\upgamma}$  = 8.0, $\boldmath{\upxi}$ = 1.5, D = 5e-9, rp = Nielsen())

# Run the NSEM simulation with the seafloor object (sf), 
# forcing dataFrame (df), and model parameters ($\uptheta$)
S=nsem(sf,df,$\uptheta$);

# Write the amplitude spectrum to a text file
writedlm("output_spectrum.txt",S)

# Set N equal to the length of the grid and and t to the length of time
N=size(S,2);
t=size(S,3);

# Correlate random noise in time with correlation coefficient, rho
u=zeros(N,N,t);
rho=0.99;

for i in 2:t
  u[:,:,i] = rho*u[:,:,i-1] + sqrt(1 - rho^2) * randn(N,N)
end

# preallocate elev
elev=zeros(N,N,t);

# Run SyntheticSeafloor on the NSEM output of the amplitude spectrum 
# time series 
for i in 1:t
    elev[:,:,i]=fourier_synthesize(sf,S[:,:,i],u[:,:,i])
end

# Write the seafloor elevation realization time series to a file
writedlm("output_elev.txt",elev)
\end{lstlisting}
%\end{verbatim}

\section{Stingray}
\label{sec:org82c9ad0}
The Stingray module implements the observation models described in Section \ref{sec:orgee79e02}. Currently, Stingray handles only the deterministic part of the observation model, leaving the full probabilistic specification of the model to be handled within the statistical inference process. The observation models are defined as types

\begin{Verbatim}[frame=single]
struct PointSpreadModel{I}
    h # The point spread function
    X # The observation locations
end
\end{Verbatim}

\noindent and 

\begin{Verbatim}[frame=single]
struct LambertianModel
    h # The measurement height
    B # The beam pattern
    R # The ranges of the observations
    theta # The azimuth angles of the observations
end
\end{Verbatim}

\noindent that can be applied to a gridded seafloor elevation field

\begin{Verbatim}[frame=single]
Y1 = (m::PointSpreadModel)(sf::Seafloor[,eta])
Y2 = (m::LambertianModel)(sf::Seafloor[,eta])
\end{Verbatim}

\noindent to obtain the deterministic response of the observation model. If an additional array \texttt{eta} is passed to these methods, the model uses \texttt{eta} as the seafloor elevation field. Otherwise, it uses the cached elevation field in the \texttt{Seafloor} object.

The \texttt{PointSpreadModel} first applies the point spread function to the gridded seafloor elevation field using a fast Fourier transform-based convolution, then interpolates to the observation locations using the interpolation method (\texttt{NearestNeighbor}, \texttt{Bilinear}, or \texttt{Bicubic}) given as the parameter \texttt{I} in the type definition. The \texttt{LambertianModel} first computes the derivatives of the gridded elevation field at the observation locations by interpolating with a kernel that is the derivative of the bicubic filter. The derivatives are then used to compute the surface normal and the surface is shaded using the Lambertian scattering model, \eqn{eq:org7365d05}. Finally, the stored beam pattern is applied to the observations, which are returned as an array with rows corresponding to the range bins and columns corresponding to the azimuth angles. Bicubic interpolation is computationally intensive when run on the CPU, so the \texttt{LambertianModel} should almost always be applied to a \texttt{Seafloor} object and an \texttt{eta} defined on the GPU.

\chapter{Architecture Demonstration}
\label{sec:org2def406}
To illustrate the application of NSEA to a real-world seafloor roughness prediction problem, the model was employed to simulate with data collected during the Target and Reverberation Experiment (TREX) off the coast of Panama City, Florida, in 2013. Instrumentation on two quadpods collected data for 34 days (April 20 -- May 23, 2013). The shallower of the quadpods was at a depth of approximately 7.5 m while the deeper was at a depth of 20 m. The present demonstration focused on data from the shallow quadpod. Wave height, period, and direction were recorded by a Nortek AWAC-AST at 2 Hz for 1,024 s every 30 minutes. Linear wave theory was used to estimate the bottom orbital velocity and the semiorbital excursion from the observed significant wave height. The forcing data input to NSEA included the bottom orbital velocity ($u$), wave semiorbital excursion ($A$), and the wave direction ($\phi$) (Fig. \ref{fig:org156906f}). A 2.25 MHz sector-scanning sonar (Imagenex 881b) installed at a height of 1.05 m above the seafloor imaged approximately a 6 m $\times$ 6 m area under the quadpods every 12 minutes. A fast Fourier transform of a 3 m $\times$ 3 m patch of the the backscatter image was computed and the wavenumber corresponding to the maximum Fourier amplitude was extracted and converted into an estimate of the ripple wavelength.

\href{figures/TREX\_forcing\_data.png}{\begin{figure}
\centering
\includegraphics[width=.9\linewidth]{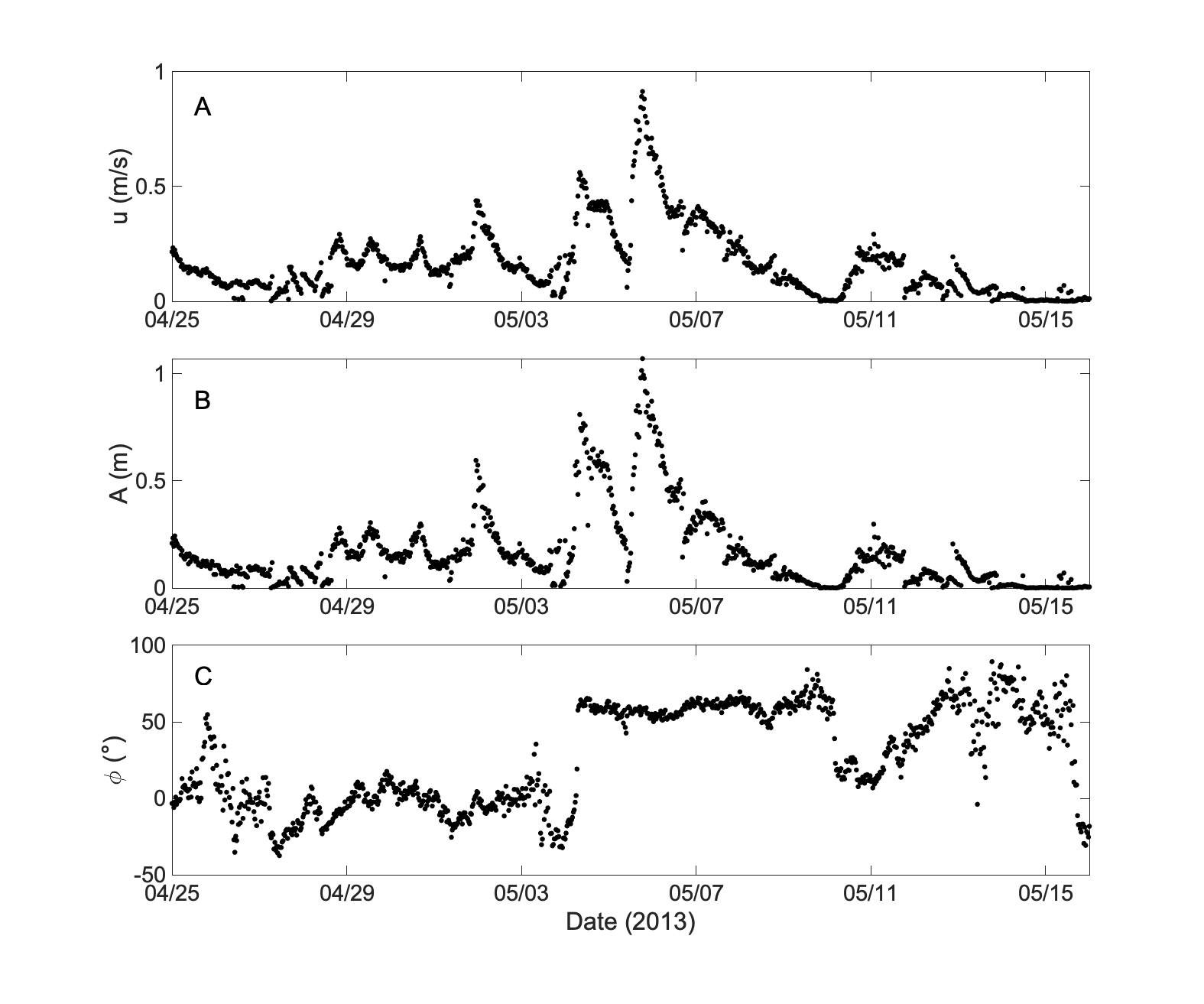}
\caption{\label{fig:org156906f} (A) Bottom orbital velocity, (B) bottom semiorbital excursion, and (C) wave direction (0$^\circ$ North) at the shallower quadpod during the TREX13 experiment}
\end{figure}}

\clearpage

\section{Forward Simulation}
\label{sec:org72414c0}
The predictive ability of NSEA was demonstrated by running the seafloor evolution model (NSEM) with the observed forcing and a set of realistic parameters given in \Tbl{tab:org30eb2e0}. The median grain size was directly measured (0.23 mm), while the other parameters were set to common values from the literature. The diffusion coefficient is set to zero; therefore, the modeled ripples do not decay in the absence of forcing. The current-generated ripple module was turned off for this run due to the negligible observed currents. NSEM simulates only wave-generated ripples using the Nelson ripple predictor \cite{Nelson_2013}. From the amplitude spectra output by NSEM, the peak wavelength was computed from the x and y wavenumbers corresponding to the maximum of the amplitude spectra. Roughness realizations were generated from the amplitude by the seafloor synthesis algorithm, and the simplified Lambertian scattering model is applied to simulate a sonar image from the roughness realizations. The range-dependent beam pattern, \(B(r)\), was estimated by computing the average over the azimuthal direction of the backscatter intensity in all of the images and dividing by the predicted backscatter response for a flat bed. The estimated beam pattern was then applied to the backscatter intensity predicted by the Lambertian scattering model so that the simulated images could be directly compared to the uncorrected sonar images from the instrument.

\begin{table}[htbp]
\caption{\label{tab:org30eb2e0}Default Parameter Values Used in the TREX NSEA Runs}
\centering
\begin{tabular}{lr}
Parameter & Value\\
\hline
\(D_{50}\) & 0.00023 m\\
S & 2.65\\
\(\theta_{cr}\) & 0.05\\
\(\theta_{wo}\) & 0.168\\
\(\alpha\) & 1.0\\
\(\varphi\) & 0.4\\
\(\sigma\) & 1.0 m\\
\(\gamma\) & 8.0\\
\(\xi\) & 1.5\\
\(D\) & 0 $\mathrm{m^{2} s^{-1}}$\\
Ripple predictor & Nelson \cite{Nelson_2013}\\
Grid size & 1024 \texttimes{} 1024\\
Patch edge length & 10 m\\
\end{tabular}
\end{table}

\begin{figure}
\centering
\includegraphics[width=.9\linewidth]{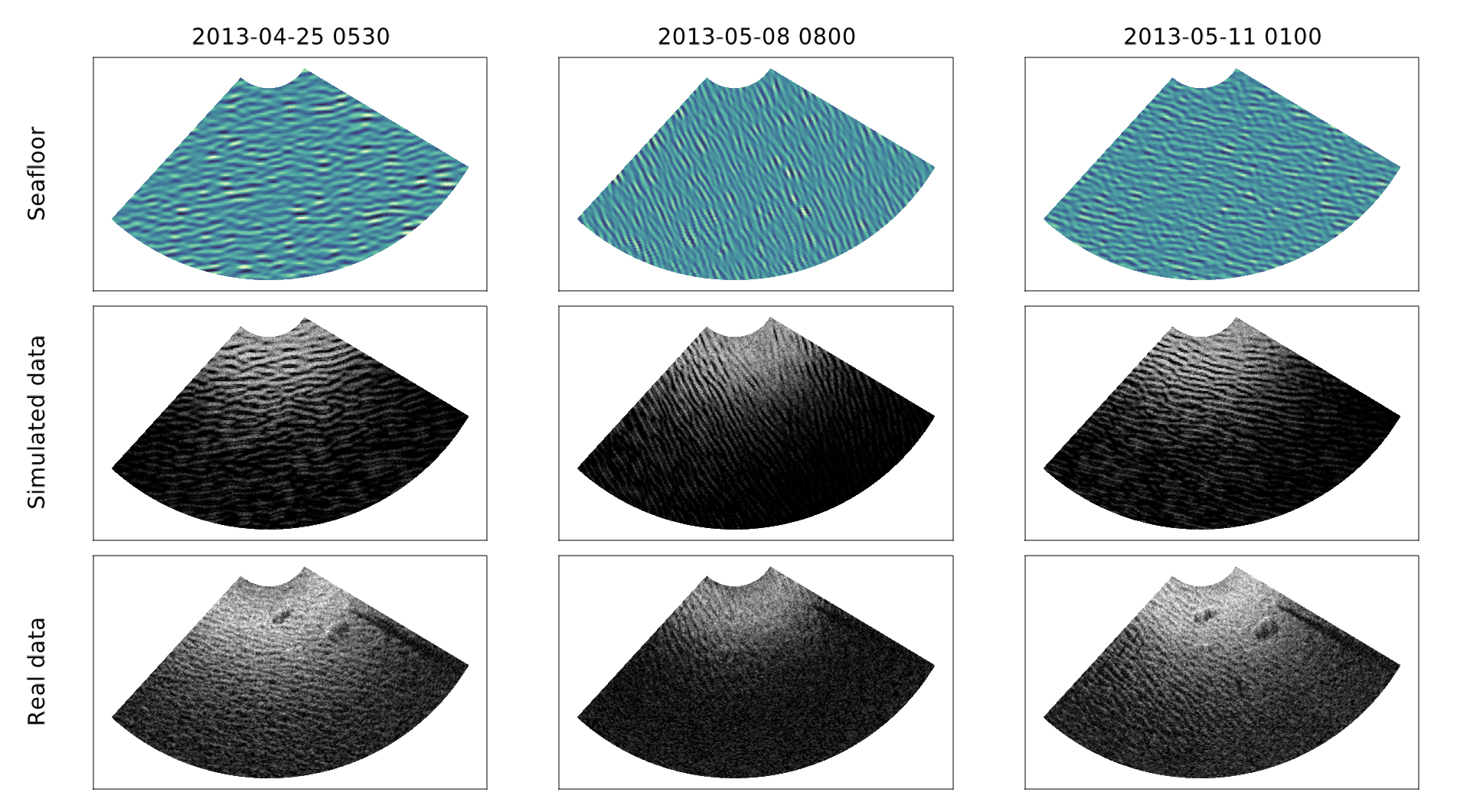}
\caption{\label{fig:org7690457}Comparison between (A) a roughness realization generated from the NSEA model output, (B) a sonar image simulated from the roughness realization, and (C) the observed sector-scanning sonar image for three different times during the TREX deployment}
\end{figure}

\begin{figure}
\centering
\includegraphics[width=.9\linewidth]{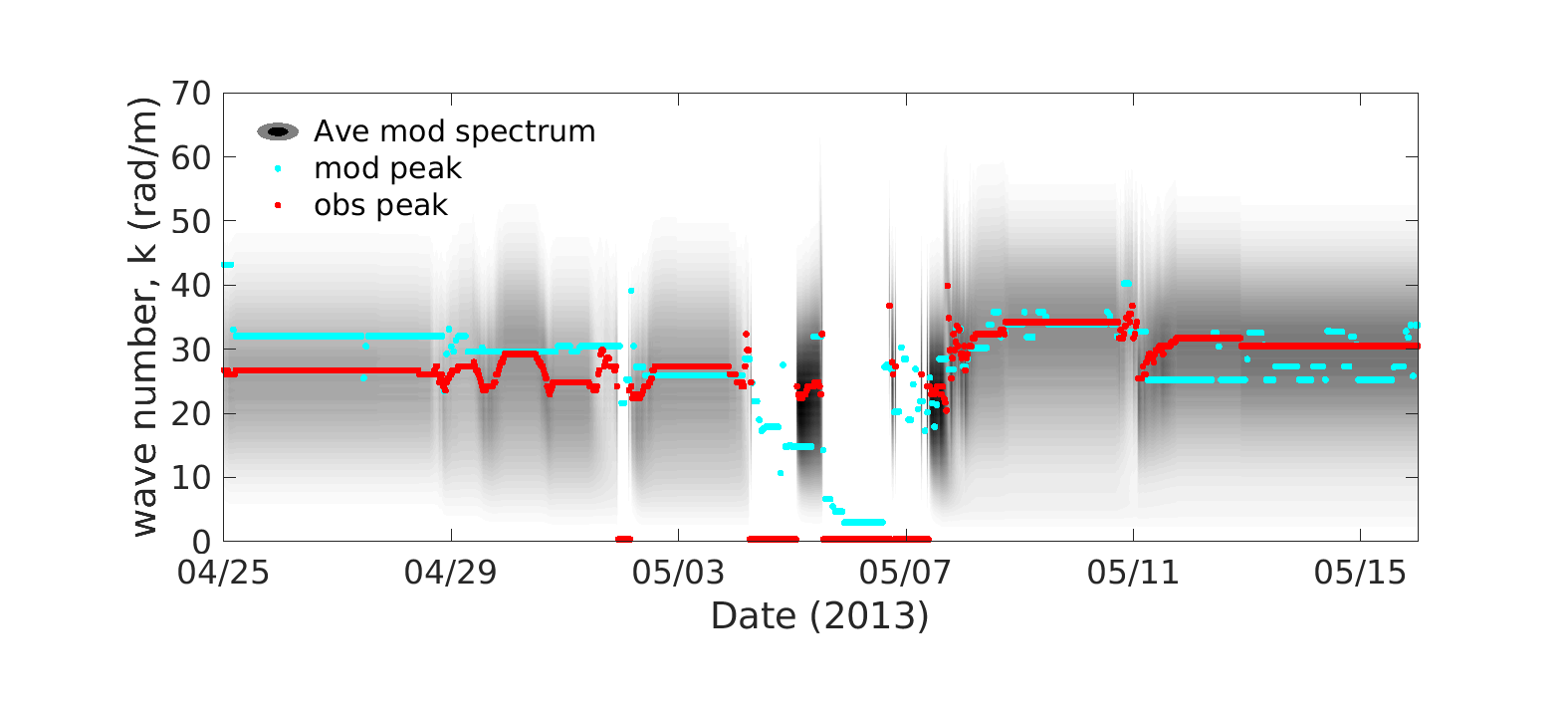}
\caption{\label{fig:org270ff7d}Comparison between the modeled (red dots) and observed (cyan dots) peak ripple wavenumbers. The grayscale contour plotted in the background represents the predicted 1D wavenumber spectrum at extracted from the peak energy direction in the 2D spectrum. The model predicts a peak wavenumber of zero when there is a washout.}
\end{figure}

The resulting roughness realizations and simulated sonar images for three times during the experiment are displayed in Fig. \ref{fig:org7690457} alongside the observed sonar images. The predicted ripple wavelengths and directions qualitatively resemble those in the observed sonar data. The roughness realizations and simulated sonar images cannot reproduce the exact pattern of roughness observed in the sonar data. In particular, the phase of the ripples in the simulated roughness realizations will never match the phase in the observed sonar data because the roughness realization is drawn randomly from the Gaussian process implied by the modeled amplitude spectrum without taking into account any of the information provided by the sonar image. Furthermore, temporal correlations between the sonar images are not accurately reproduced by the model because the roughness realizations are independently drawn at each time step. The peak ripple wavelength time series predicted by the model is compared to the observed ripple wavelengths extracted from the imagery by the FFT in Fig. \ref{fig:org270ff7d}. Again, the model predictions agree fairly well with the data. The discrepancy between the model and the observations at the beginning of the time series exists because NSEM initializes the seafloor spectrum to be in equilibrium with the hydrodynamic forcing, even if the forcing is not strong enough to mobilize sediment and to generate ripples. In the case of the TREX data, long-wavelength relict ripples were present at the site at the beginning of the deployment. These ripples were formed by stronger forcing some time prior to the deployment and existed out of equilibrium with the hydrodynamics. Since NSEM does not have any information about the forcing prior to the deployment, it is unable to predict the geometry of the relict ripples, and the discrepancy persisted until the first sediment mobilization event around April 24, 2013. Once sediment was mobilized, the seafloor ripples reset, and the evolution model predicted a new ripple wavelength within a few centimeters of the observations. The end of the time series illustrates the ability of NSEM to predict the persistence of relict ripples that formed during its simulation period.
\clearpage

\section{Statistical Inference}
\label{sec:org1d9f9a2}

The parameters of the seafloor evolution model can be estimated from the TREX forcing data and the ripple wavelength data using approximate Bayesian computation. The prior distribution for the parameters was chosen to be uniform with lower and upper bounds selected by expert elicitation and shown in \Tbl{tab:org7a0db52}. The prior distribution of the power law ripple predictor, Eqs. \eqref{eq:org13961b3} and \eqref{eq:org13961b3a}, parameters was determined by fitting the power law model to the compiled ripple data from Nelson et al. \cite{Nelson_2013}. One million parameter samples were taken from the prior distribution and were run through the model using the forcing data shown in Fig. \ref{fig:org156906f}. The peak wavelength at each time step was then computed from the wavenumber corresponding to the maximum of the amplitude spectrum. The discrepancy measure used was the mean squared deviation between the predicted and the observed ripple wavelength, and the tolerance was chosen to accept approximately 1,000 parameter samples, based on the acceptance rate of a pilot run.
\clearpage

\begin{table}%[htbp]
\caption{\label{tab:org7a0db52}Lower and Upper Bounds of the Uniform Prior Distribution for Each of the Parameters}
\centering
\begin{tabular}{lrr}
Parameter & Lower bound & Upper bound\\
\hline
\(D_{50}\) & 0.0001 mm & 0.002 mm\\
S & 2.5 & 2.8\\
\(\theta_{cr}\) & 0.01 & 0.07\\
\(\theta_{wo}\) & 0.01 & 1.0\\
\(\alpha\) & 1.0 & 20.0\\
\(\varphi\) & 0.3 & 0.45\\
\(\sigma\) & 1.0 m & 25.0 m\\
\(\gamma\) & 1.0 & 15.0\\
\(\xi\) & 0.0 & 2.5\\
\(D\) & 0 $\mathrm{m^{2} s^{-1}}$ & 1 \texttimes{} 10\(^{\text{-12}}\) $\mathrm{m^{2} s^{-1}}$\\
\(\beta_1\) & -1.4 & -1.0\\
\(\beta_2\) & -0.2 & 0.1\\
\(\beta_3\) & -0.2 & 0.1\\
\(\chi_1\) & 1.7 & 2.2\\
\(\chi_2\) & -0.6 & -0.3\\
\(\chi_3\) & -0.2 & 0.0\\
\end{tabular}
\end{table}

\begin{figure}
\centering
\includegraphics[width=.9\linewidth]{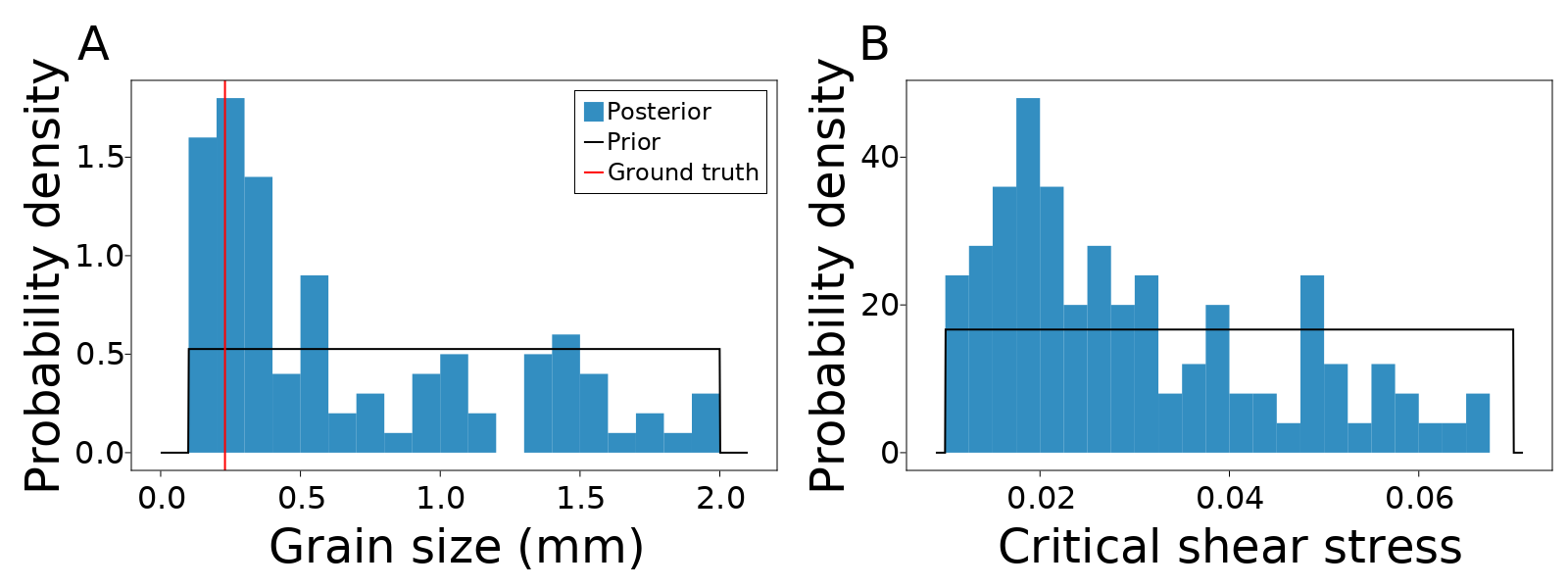}
\caption{\label{fig:orgaf15a14}Approximate posterior distributions for the median grain size (A) and the critical shear stress (B). The uniform prior distribution is indicated with the black line. The ground truth grain size is indicated with the red line in plot (A).}
\end{figure}

The posterior distributions for the median grain size and the critical shear stress are shown in Fig. \ref{fig:orgaf15a14}. The peak of the posterior distribution corresponds to the most likely parameter value in the model, while the concentration of the posterior distribution relative to the prior distribution (shown with a black line in Fig. \ref{fig:orgaf15a14}) indicates how sensitive the model is to that parameter. The grain size posterior peaks around a value close to the ground truth of 0.23 mm and concentrates relative to the prior. The seafloor evolution model's sensitivity to the grain size is to be expected, since the grain size relates the scales of the flow to the bed response and thus appears as a scaling parameter in many of the empirical formulae used in the seafloor evolution model, such as the bedload transport law (Eq. \eqref{eq:org0795cab}) and the ripple predictors (Eqs. \eqref{eq:org13961b3} and \eqref{eq:org13961b3a}). The critical shear stress posterior concentrates around a value of 0.02. While there were no independent measurements of critical shear stress at this site, the estimated value is lower than is typically used in sediment transport modeling studies ($\sim0.05$ for quartz sand). The critical shear stress sets the timing of sediment mobilization and the magnitude of sediment transport but does not play a role in the ripple geometry; therefore, the lower estimated critical shear stress indicates that the ripples started evolving sooner and evolved more quickly than has been historically observed. This discrepancy could indicate the importance of presently unmodeled processes in the timing of sediment mobilization; however, it is also possible that the inference algorithm uses the flexibility in the critical shear stress parameter to overcome some other discrepancy between the model and the data. More sophisticated sensitivity analyses are necessary to discriminate between these possibilities and to improve both the model and the inference procedure \cite{Gelman_2014}.
\clearpage

\chapter{Conclusions}
\label{sec:orgf3071a0}
The Naval Seafloor Evolution Architecture is a flexible platform for forecasting seafloor roughness. NSEA is based on a simplified sediment transport model that predicts the statistical properties of the seafloor elevation under changing hydrodynamic conditions. It provides a Bayesian inference framework to combine the seafloor evolution model with hydrodynamic forcing information and observations from diverse sources. NSEA can be run in forward mode to predict seafloor roughness from wave and current forcing and to quantify the uncertainty in those predictions. Downstream applications such as acoustic or hydrodynamic models can use these roughness forecasts as boundary conditions. NSEA can also be run in an inverse mode using observations to estimate roughness or the parameters of the seafloor evolution model. The Bayesian approach unifies these two formulations into the single task of estimating the conditional probability distribution of the quantity of interest given the available information.

NSEA's modular architecture separates the specification of hydrodynamic forcing, the seafloor evolution model, and the generation of roughness spectra, seafloor elevation, and acoustic sonar response predictions. For example, the seafloor evolution model is agnostic to the source of the hydrodynamic forcing information, so the forcing can be provided from in situ observations, as illustrated in \Secn{sec:org2def406}, or from hydrodynamic models. The seafloor evolution model could also be used independently to feed roughness information back into a hydrodynamic model. Finally, the acoustic observation model can be substituted to take advantage of the complementary information available from different observation modalities. An example of a simplified acoustic model for high-frequency imaging sonars is described in \Secn{sec:org854c271}. However, with more sophisticated acoustic models, it is possible to constrain seafloor roughness and the parameters of the seafloor evolution model using many different kinds of acoustic observations. NSEA's modularity means that such models, developed by independent teams, can be combined with the seafloor evolution model and Bayesian inference tools with minimal effort.

\section{Future Work}
\label{sec:org1f95028}
NSEA was originally developed to model the evolution of roughness due to wave-generated ripples on shallow, sandy seafloors, and it has been extensively tested in such environments. For NSEA to realize the goal of producing forecasts of seafloor roughness globally requires additional research and validation. NSEA must be extended to consider the modification of roughness by processes other than surface waves. Current-generated ripples can be simulated by NSEA, but are less extensively tested, and the wave and current ripple parameterizations do not currently simulate the complex patterns that can be generated by combined wave and current flows. Larger-scale bedforms can also be created by currents. While there are few theoretical limitations in simulating large-scale roughness with NSEA's spectral approach, new empirical parameterizations need to be incorporated for NSEA to generate realistic bedforms at larger scales. In deeper water, surface waves have less influence on the bottom and processes such as geostrophic currents, internal waves and turbidity currents become more important. These can generate a large variety of geomorphological features beyond simple bedforms. Representing these kinds of forcings and their impacts on the seafloor is essential to extend the applicability of NSEA into deeper waters. Applying NSEA globally also requires mechanisms for predicting roughness and its evolution in mud, silt, and gravel sediment and in hard-bottom environments like bedrock outcrops and coral reefs.

Biological processes can also modify roughness. Organisms moving and feeding on the bottom can create and destroy roughness. In addition, the growth of sediment-associated organisms like microphytobenthos can affect the mobility of sediment and therefore the evolution of bedforms. At present, NSEA represents biological processes as a diffusion coefficient that smoothes roughness during quiescent conditions. Even this simplified model poses challenges of estimating the diffusion coefficient in different locations and times of year. To properly represent the effect of biological activity on roughness evolution likely requires coupling NSEA to ecological models. 

The major challenge of all of these extensions is to ensure that the complexity required to represent these different processes is justified by increased predictive power. Each new process adds parameters that need to be calibrated, which can increase the overall uncertainty in the model's predictions even if the new processes make those predictions more accurate. An important benefit of framing NSEA as a probabilistic model is the ability to evaluate the model against observations and to make rigorous statistical comparisons between different formulations of the model. As NSEA is further developed and applied to new settings in the future, it should constantly be tested against observations to ensure that new parameterizations create better predictions of roughness evolution.

\bibliography{library}
\end{document}